\documentclass[twocolumn,aps,prd,amsmath,amssymb,preprintnumbers,nofootinbib]{revtex4-1}

\usepackage{graphicx,caption}
\usepackage{tensor}
\usepackage{xcolor}
\usepackage{braket}
\usepackage{multirow}

\begin{document}

\setcounter{footnote}{0}

\title{Hamiltonian gravity with fermions}
\author{Erick I.\ Duque}
\email{eqd5272@psu.edu}
\affiliation{Institute for Gravitation and the Cosmos,
The Pennsylvania State University, 104 Davey Lab, University Park,
PA 16802, USA}

\begin{abstract}
Fermions are coupled to the Einstein--Cartan system in the canonical formulation, including the cosmological, the Barbero--Immirzi, and the non-minimal coupling constants.
The resulting ten first-class constraints generate gauge transformations that are on-shell equivalent to spacetime diffeomorphisms and SL(2,$\mathbb{C}$) transformations.
The gravitational second-class constraints receive fermionic contributions, which can be implemented by use of Dirac brackets or by solving them directly.
Furthermore, we identify new fermionic (second-class) constraints that are necessary to recover the Dirac-fermion theory by relating the momenta to the configuration variables on dynamical solutions; this fermionic phase-space reduction is accomplished by use of corresponding Dirac brackets.
The theory remains well-defined off the second-class constraint surfaces with ten additional degrees of freedom---six of which are gravitational and the remaining four are fermionic.
Discrete ($CPT$) symmetries as well as implications for canonical quantization and modified theories of gravity with fermions are discussed.
\end{abstract}

\maketitle

\section{Introduction}

Fermions are notoriously complicated to manipulate algebraically.
This is the main reason why most studies of quantum field theory focus on scalar fields instead and why the Hamiltonian formulation of fermions coupled to gravity, even in the classical context, remains largely unexplored.

In Lagrangian formulations, it is relatively straightforward to couple fermions to gravity in the Einstein--Cartan theory, which uses tetrads and connection variables, implying a larger gravitational field content compared to Einstein's metric system.
Canonical treatments of the Einstein--Cartan theory in vacuum, based on the Holst action (given by the Hilbert--Palatini action with a nontrivial Barbero--Immirzi constant), usually work in Ashtekar--Barbero variables \cite{Ashtekar1,Ashtekar2,Barbero,Holst}, which, however, fix a particular gauge from the start and hence break the underlying Lorentz invariance of the action.
Therefore, any canonical quantum gravity theory based on such a system is not guaranteed to be Lorentz covariant.
This problem was resolved in \cite{Tetrads} by performing the canonical decomposition of the Holst action without gauge fixing.
This procedure results in a larger phase space compared to Ashtekar--Barbero variables and the resulting constraints generate gauge transformations that are on-shell equivalent to spacetime diffeomorphisms and SO(1,3) transformations, contrary to the Ashtekar--Barbero system which cannot generate boosts.
The main purpose of this paper is to couple fermions in this new formulation to preserve the full SL(2,$\mathbb{C}$) transformations of the Einstein--Cartan theory; this is a necessary step preceding canonical approaches to diffeomorphism and Lorentz-covariant theories of quantum or modified gravity with fermionic matter.

This work is organized as follows.
We start with a brief review of the Lagrangian formulation of the Einstein--Cartan system coupled to a massless Dirac fermion, including the cosmological, Barbero--Immirzi, and non-minimal coupling constants in Section~\ref{sec:Lagrangian formulation}.
In Section~\ref{sec:Canonical formulation}, the canonical analysis of this system is performed, identifying the symplectic structure, the non-dynamical variables, and the decomposition of useful spacetime quantities in terms of the variables of the extended phase space; we also identify hitherto neglected fermionic constraints that are necessary to relate the extended phase-space to that of the Dirac fermions via a phase-space reduction.
In Section~\ref{sec:First-class constraints}, we compute the first-class constraints and their algebra under Poisson brackets of the extended phase space, study the gauge transformations of the system and their relation to spacetime diffeomorphisms and SL(2,$\mathbb{C}$) transformations, and identify a set of (nonlocal) Dirac observables that imply conserved spin and electric currents.
In Section~\ref{sec:Second-class constraints}, we obtain the second-class constraints of the theory necessary to reduce the extended phase space to that implied by the action and derive the corresponding Dirac brackets.
We incorporate the mass term in Section~\ref{sec:Mass term}.
In Section~\ref{sec:Canonical quantization}, we address some misconceptions about the canonical quantization of fermionic variables that led to some commutation inconsistencies \cite{Inconsistency,Half-densitized}, which are resolved by a proper understanding of the phase-space reduction that is necessary to describe Dirac fermions in canonical formulations.
In Section~\ref{sec:Discrete symmetries}, we study the system's behavior under charge conjugation ($C$), parity ($P$), and time-reversal ($T$) transformations.
In Section~\ref{sec:EMGFT}, we outline future applications to (emergent) modified gravity theories.
The conclusions are presented in Section~\ref{sec:Conclusions}.

We work exclusively in four dimensions with Lorentzian signature and set $c=1$.
Greek letters are used for spacetime indices and capital Latin letters $I,J,K,.\,.\,. \in \{0,1,2,3\}$ for internal Lorentz indices.
We use the Latin letters $a,b,c,\dots,h$ for spatial indices, and $i,j,k,.\,.\,.\in\{1,2,3\}$ will be reserved for internal Euclidean indices.
We use dotted capital Latin letters $\dot{A},\dot{B},\dots$ for the Dirac spinor index, which will usually be suppressed but its use facilitates the handling of Poisson brackets later; similarly, the dotted Latin letters $\dot{a},\dot{b},\dots\in\{1,2\}$ are used for left-handed spinor indices, and double-dotted Latin letters $\ddot{a},\ddot{b},\dots\in\{1,2\}$ are used for right-handed spinor indices.

\section{Lagrangian formulation}
\label{sec:Lagrangian formulation}

\subsection{Kinematics}

Spacetime geometry can be described by a tetrad $e^\mu_I$ such that the metric is given by the inverse of
\begin{equation}\label{eq:Inverse metric in tetrads}
    g^{\mu\nu}=\eta^{IJ}e_I^\mu e_J^\nu\,,
\end{equation}
where $\eta_{IJ}$ is the Minkowski metric.
Spacetime indices are raised and lowered with $g_{\mu\nu}$ and internal Lorentz indices with $\eta_{IJ}$.
Relation (\ref{eq:Inverse metric in tetrads}) implies
\begin{equation}
    g_{\mu\nu} e^\mu_I e^\nu_J = \eta_{IJ}\,.
\end{equation}
This spacetime metric defines the derivative operator $\nabla_\mu$ compatible with it,
\begin{eqnarray}
    \nabla_\alpha g_{\mu\nu}=0\,,
\end{eqnarray}
which is used to parallel transport spacetime tensors.

On the other hand, a connection 1-form $\tensor{\omega}{_\mu^I^J}$ is necessary to parallel transport SO(1,3)-valued tensor fields $f^{I_1,I_2,\dots,I_n}$ with the covariant derivative
\begin{equation}
    D_\mu f^{I_1,\dots,I_n} = \nabla_\mu f^{I_1,\dots,I_n} + \sum_{k=1}^n \tensor{\omega}{_\mu^{I_k}_J} f^{I_1,\dots,J,\dots,I_n}\,,
\end{equation}
The antisymmetry of the connection implies that the internal Minkowski metric remains invariant under parallel transport,
\begin{equation}
    D_\mu \eta_{IJ}=0\,.
\end{equation}

A general spacetime geometry has torsion, defined as
\begin{eqnarray}\label{eq:Torsion}
    \tensor{T}{^I_\mu_\nu} \!\!&\equiv&\!\! D_{[\mu} e_{\nu]}^I
    \nonumber\\
    \!\!&=&\!\! \partial_{[\mu} e_{\nu]}^I
    + \tensor{\omega}{_{[\mu}^I_{|K|}} e^K_{\nu]}\,.
\end{eqnarray}

Consider a Dirac spinor
\begin{equation}
    \Psi^{\dot{A}}=\left(\begin{matrix}
        \psi_{\dot{a}}\\
        \chi^{\ddot{a}}
    \end{matrix}\right)\,,
\end{equation}
where $\psi$ is a left-handed Weyl field and $\chi$ is right-handed.
The complex-conjugate of the Dirac spinor is given by $\overline{\Psi}=\Psi^\dagger\beta$.
(Complex conjugation changes left to right-handed indices and viceversa: $(\psi_{\dot{a}})^*=\psi^*_{\ddot{a}}$, $(\psi^{\ddot{a}})^*=\psi^{*\dot{a}}$.)
Here,
\begin{equation}
    \beta = \left(\begin{matrix}
        0 & \tensor{\delta}{^{\ddot{a}}_{\ddot{b}}}\\
        \tensor{\delta}{_{\dot{a}}^{\dot{b}}} & 0
    \end{matrix}\right)\,.
\end{equation}
We introduce the invariant symbols $\varepsilon_{\dot{a}\dot{b}}$, $\varepsilon^{\dot{a}\dot{b}}$, $\varepsilon_{\ddot{a}\ddot{b}}$, and $\varepsilon^{\ddot{a}\ddot{b}}$ such that
\begin{eqnarray}\label{eq:Invariant spinor symbol}
    \varepsilon^{\dot{1}\dot{2}} = \varepsilon^{\ddot{1}\ddot{2}}
    = \varepsilon_{\dot{2}\dot{1}} = \varepsilon_{\ddot{2}\ddot{1}}
    = + 1
    \,,\nonumber\\
    \varepsilon^{\dot{2}\dot{1}} = \varepsilon^{\ddot{2}\ddot{1}}
    = \varepsilon_{\dot{1}\dot{2}} = \varepsilon_{\ddot{1}\ddot{2}}
    = - 1
    \,;
\end{eqnarray}
it follows that $\varepsilon_{\dot{a}\dot{b}}\varepsilon^{\dot{b}\dot{c}}=\tensor{\delta}{_{\dot{a}}^{\dot{c}}}$ and $\varepsilon^{\dot{a}\dot{b}}\varepsilon_{\dot{b}\dot{c}}=\tensor{\delta}{^{\dot{a}}_{\dot{c}}}$.
We use (\ref{eq:Invariant spinor symbol}) to raise and lower spinorial indices: $\psi^{\dot{a}}=\varepsilon^{\dot{a}\dot{b}}\psi_{\dot{b}}$, $\chi_{\ddot{a}}=\varepsilon_{\ddot{a}\ddot{b}}\chi^{\ddot{b}}$.
In the following, spinorial indices will be suppressed when no confusion arises and assume that whenever two spinorial tensors appear together, their proximal indices are contracted according to the following convention: The contraction between two left-handed spinors is given by $\psi\xi=\psi^{\dot{a}}\xi_{\dot{a}}$, between two right-handed spinors by $\xi^\dagger\chi^\dagger=\xi^\dagger_{\ddot{a}}\chi^{\dagger\ddot{a}}$, between left and right-handed spinors by $\psi^\dagger \psi=\psi^{\dagger}{}^{\ddot{a}} \tensor{\delta}{_{\ddot{a}}^{\dot{a}}} \psi_{\dot{a}}$, and similarly for other spinorial tensors.

The action of the covariant derivative on a Dirac spinor is given by
\begin{equation}
    D_\mu \Psi = \partial_\mu \Psi + \frac{i}{2} \tensor{\omega}{_{\mu}^I^J} S_{IJ} \Psi
    \,,
\end{equation}
where
\begin{equation}
     S^{IJ} = \frac{i}{4} \left[\gamma^I,\gamma^J\right]\,,
\end{equation}
and
\begin{equation}
    \gamma^I= \left(\begin{matrix}
        0 & \sigma^I\\
        \bar{\sigma}^I & 0
    \end{matrix}\right)
\end{equation}
are the Dirac matrices---not to be confused with the boosting function $\gamma=1/\sqrt{1-v^2}$, which has no indices, or $\gamma^2$, which will always refer to the squared boosting function here---with
\begin{equation}
    \sigma^I_{\dot{a}\ddot{a}} = (\delta_{\dot{a}\ddot{a}},\sigma^i_{\dot{a}\ddot{a}})
    \quad,\quad
    \bar{\sigma}^{I\ddot{a}\dot{a}} = (\delta^{\ddot{a}\dot{a}},-\sigma^{i\ddot{a}\dot{a}})\,,
\end{equation}
where $\sigma^i$ are the Pauli matrices.
The covariant derivative of the Dirac co-spinor is given by
\begin{equation}
    D_\mu \overline{\Psi} = \partial_\mu \overline{\Psi} - \frac{i}{2} \tensor{\omega}{_{\mu}^I^J} \overline{\Psi} S_{IJ}
    \,,
\end{equation}
such that $D_\mu \overline{\Psi}=\overline{D_\mu \Psi}$ holds.

\subsection{Dynamics}

The vacuum dynamics are generated by the action \cite{Holst}
\begin{eqnarray}\label{eq:Holst action}
    S_G[e,\omega] \!\!&=&\!\! \int {\rm d}^4x \frac{|e|}{16\pi G} \left[e^\mu_I e^\nu_J \tensor{P}{^I^J_K_L} F^{KL}_{\mu\nu} - 2 \Lambda\right]
    \,,\nonumber
\end{eqnarray}
where $\Lambda$ is the cosmological constant, $|e|$ is the absolute value of the determinant of the co-tetrad $e_\mu^I$, $\epsilon_{IJKL}$ is a totally antisymmetric internal tensor with $\epsilon_{0123}=-1$,
\begin{equation}
    F^{IJ}_{\mu\nu} = 2 \partial_{[\mu} \omega_{\nu]}^{IJ}
    + 2 \omega_{[\mu}^{IK} \omega_{\nu]}^{LJ} \eta_{KL}
\end{equation}
is the strength tensor field of the connection, and
\begin{equation}
    \tensor{P}{^I^J_K_L} = \delta^{[I}_K \delta^{J]}_L
    - \frac{\zeta}{2} \tensor{\epsilon}{^I^J_K_L}\,,
\end{equation}
where $\zeta\in\mathbb{R}$ is the inverse of the Barbero--Immirzi parameter.
The last expression has the inverse
\begin{equation}
    \tensor{(P^{-1})}{^K^L_I_J} = \frac{1}{1+\zeta^2} \left(\delta^{[K}_I \delta^{L]}_J
    + \frac{\zeta}{2} \tensor{\epsilon}{^K^L_I_J}\right)
    \,,
\end{equation}
such that $\tensor{P}{^I^J_M_N}\tensor{(P^{-1})}{^M^N_K_L}=\delta^{[I}_K\delta^{J]}_L$.

In the presence of a massless Dirac fermion, the full action $S=S_G+S_F$ receives a fermionic contribution given by \cite{Mercuri,Perez,Freidel,Alexandrov}
\begin{widetext}
\begin{eqnarray}\label{eq:Dirac action}
    S_F \!\!&=&\!\! \int {\rm d}^4x |e| \frac{i}{2} e^\mu_I \left(\overline{\Psi} \gamma^I \left(1-i\alpha\gamma^5\right) D_\mu \Psi
    - D_\mu \overline{\Psi} \left(1-i\alpha\gamma^5\right) \gamma^I \Psi\right)
    \\
    \!\!&=&\!\! \int {\rm d}^4x |e| \bigg[ \frac{i}{2} e^\mu_I \left(\overline{\Psi} \gamma^I \left(1-i\alpha\gamma^5\right) \partial_\mu \Psi
    - \partial_\mu \overline{\Psi} \left(1-i\alpha\gamma^5\right) \gamma^I \Psi\right)
    - \frac{1}{4} e^\mu_K \tensor{\omega}{_{\mu}^I^J} \overline{\Psi} \left([\gamma^K,S_{IJ}]_++i\alpha\gamma^5[\gamma^K,S_{IJ}]\right) \Psi \bigg]
    \nonumber\\
    \!\!&=&\!\! \int {\rm d}^4x |e| \bigg[ \frac{i}{2} e^\mu_I \left(\overline{\Psi} \gamma^I \left(1-i\alpha\gamma^5\right) \partial_\mu \Psi
    - \partial_\mu \overline{\Psi} \left(1-i\alpha\gamma^5\right) \gamma^I \Psi\right)
    + \frac{1}{4} e^\mu_K \tensor{\omega}{_{\mu}^I^J} \overline{\Psi} \left(\tensor{\epsilon}{^K^L_I_J} + 2 \alpha \delta^K_I \delta^L_J \right) \gamma_L \gamma^5 \Psi\bigg]
    \,,\nonumber
\end{eqnarray}
\end{widetext}
where $\alpha\in\mathbb{R}$ is the non-minimal coupling parameter, $\gamma^5=- \frac{i}{4!} \epsilon_{IJKL} \gamma^I\gamma^J\gamma^K\gamma^L$, and we used
\begin{equation}\label{eq:[gamma,S]+}
    \left[\gamma^I,S^{JK}\right]_+
    = - \tensor{\epsilon}{^I^J^K_L} \gamma^L \gamma^5\,,
\end{equation}
\begin{equation}
    \left[\gamma^I,S^{JK}\right]
    = - 2 i \eta^{I[J} \gamma^{K]}
    \label{eq:[gamma,S]-}
\end{equation}
to get the last line.

Using $\delta F^{IJ}_{\mu\nu}=2 D_{[\mu} \delta \omega^{KL}_{\nu]}$, the variation of the action with respect to the connection 1-form, upon some simplification and neglecting boundary terms, is given by
\begin{eqnarray}\label{eq:Torsion-constraints}
    e_\mu^P \frac{\delta S_G}{\delta \omega_\mu^{IJ}}
    \!\!&=&\!\! e_\mu^P \frac{\epsilon^{\mu\nu\alpha\beta}}{32\pi G} \tensor{P}{_I_J^K^L} \epsilon_{KLMN} D_\nu \left(e^M_\alpha e^N_\beta\right)
    \nonumber\\
    \!\!&=&\!\!
    - 3! \frac{|e|}{16\pi G} \tensor{P}{_I_J^K^L} \delta^{P}_{[K} \tensor{T}{^N_L_{N]}}\,,
\end{eqnarray}
where $\tensor{T}{^I_J_K}=e^\mu_Je^\nu_K\tensor{T}{^I_\mu_\nu}$, from the gravitational contribution, and
\begin{equation}\label{eq:Torsion-constraints - fermionic}
    e_\mu^P \frac{\delta S_F}{\delta \omega_\mu^{IJ}}
    = - \frac{|e|}{4} \overline{\Psi} \left([\gamma^P,S_{IJ}]_++i\alpha\gamma^5[\gamma^P,S_{IJ}]\right) \Psi
\end{equation}
from the fermionic contribution.
Therefore, the equation of motion $\delta S/\delta \omega_\mu^{IJ}=0$ can be written as the vanishing of
\begin{widetext}
\begin{eqnarray}\label{eq:Torsion EoM}
    \tensor{{\cal T}}{^P_\mu_\nu} \!\!&\equiv&\!\! - \frac{8\pi G}{|e|} e_\alpha^P e^I_\mu e^J_\nu \tensor{(P^{-1})}{_I_J^K^L} \frac{\delta S}{\delta \omega_\alpha^{KL}}
    \\
    \!\!&=&\!\! e^I_\mu e^J_\nu \left[3 \delta^{P}_{[I} \tensor{T}{^N_J_{N]}}
    + \frac{8\pi G}{4} \tensor{(P^{-1})}{_I_J^R^S} \overline{\Psi} \left([\gamma^P,S_{RS}]_++i\alpha\gamma^5[\gamma^P,S_{RS}]\right) \Psi\right]\,.
    \nonumber
\end{eqnarray}
In particular, notice that this equation of motion implies
\begin{eqnarray}
    \tensor{T}{^N_I_N}
    \!\!&=&\!\!
    \frac{1}{2} \frac{8\pi G}{4} \tensor{(P^{-1})}{_I_N^R^S} \overline{\Psi} \left([\gamma^N,S_{RS}]_++i\alpha\gamma^5[\gamma^N,S_{RS}]\right) \Psi
    \nonumber\\
    \!\!&=&\!\!
    \frac{1}{2} \frac{8\pi G}{4} \frac{1}{1+\zeta^2} \overline{\Psi} \left(\frac{\zeta}{2} \tensor{\epsilon}{_I_N^R^S} [\gamma^N,S_{RS}]_++i\alpha\gamma^5[\gamma^N,S_{IN}]\right) \Psi
    \,.
\end{eqnarray}
\end{widetext}
Using
\begin{equation}
    |e| e^\nu_Ie^\mu_JT^I_{\mu\nu}=\frac{|e|}{2}D_\nu e^\nu_J\,,
\end{equation}
we conclude that, in the presence of fermionic matter and nontrivial Barbero--Immirzi or non-minimal coupling parameters, the covariant divergence of the tetrad is non-vanishing:
\begin{eqnarray}\label{eq:Divergence densitized tetrad}
    D_\mu \left(|e|e^\mu_I\right) \!\!&=&\!\! \frac{8\pi G}{4} \frac{|e|}{1+\zeta^2} \overline{\Psi} \bigg(\frac{\zeta}{2} \tensor{\epsilon}{_I_N^R^S} [\gamma^N,S_{RS}]_+
    \nonumber\\
    \!\!&&\!\!\qquad\qquad\qquad\quad
    +i\alpha\gamma^5[\gamma^N,S_{IN}]\bigg) \Psi
    \nonumber\\
    \!\!&=&\!\! - \frac{3}{4} \frac{8\pi G |e|}{1+\zeta^2} \overline{\Psi} \left(\zeta-\alpha\right) \gamma_I \gamma^5 \Psi
    \,.
\end{eqnarray}

On the other hand, the fermionic equation of motion $\delta S/\delta \overline{\Psi}=0$ is given by
\begin{equation}\label{eq:Dirac equation}
    i |e| e^\mu_I \gamma^I D_\mu \Psi
    + \frac{i}{2} D_\mu \left(|e| e^\mu_I\right) \left(1-i\alpha\gamma^5\right) \gamma^I \Psi = 0
    \,.
\end{equation}
The second term in this equation is generally non-trivial according to (\ref{eq:Divergence densitized tetrad}).
While $\Psi$ and $\overline{\Psi}$ are kinematically (off shell) independent in the action (\ref{eq:Dirac action}), the equation of motion $\delta S/\delta \Psi=0$ yielding the complex-conjugated version of (\ref{eq:Dirac equation}) guarantees that, on dynamical solutions (on shell), the reality condition $\overline{\Psi}=\Psi^\dagger \beta$, such that $\overline{\Psi}$ is indeed related to $\Psi$ by complex conjugation, can be consistently imposed.

\subsection{Degrees of freedom}

The 24 equations of motion derived from $\delta S/\tensor{\omega}{_\mu^I^J}=0$ can be seen as equations for the torsion $T_{\mu\nu}^I$ with a source and hence correspond to only 12 independent equations with a first-order time derivative of $e^\mu_I$, while the remaining 12 equations contain only spatial derivatives.
Therefore, the latter are not evolution equations but constraints, implying that the connection has only 12 dynamical components.
Similarly, from the 16 equations of motion derived from $\delta S/\delta e^\mu_I=0$, only 12 contain first-order time derivatives of the connection, while the other 4 components contain only spatial derivatives.
Therefore, the latter are constraints too and the tetrad has only 12 dynamical components.
From the total of 16 constraints, 6 are second class and other 10 are first class \cite{Tetrads}.
This counting of constraints implies that the 12 pairs of dynamical gravitational field components reduce to expected 2 degrees of freedom on dynamical solutions---each first-class constraint determines a pair of dynamical field components.

On the other hand, the total of 16 components of the kinematically independent fermion fields $\Psi$ and $\overline{\Psi}$ reduce to only 8 upon imposing the reality conditions $\overline{\Psi}=\Psi^\dagger \beta$.
Furthermore, because the resulting Dirac equation (\ref{eq:Dirac equation}) is a first-order differential equation, the 8 components of $\Psi$ reduce to only 4 degrees of freedom on dynamical solutions.

\section{Canonical formulation}
\label{sec:Canonical formulation}

\subsection{Foliation}
\label{sec:Foliation}

Given a hyperbolic spacetime region, $M = \Sigma \times \mathbb{R}$, the line element in ADM form is given by \cite{ADM,arnowitt2008republication}
\begin{equation}
  {\rm d} s^2 = - N^2 {\rm d} t^2 + q_{a b} ( {\rm d} x^a + N^a {\rm d} t )
  ( {\rm d} x^b + N^b {\rm d} t )
  \,,
  \label{eq:ADM line element}
\end{equation}
where $N$ is the lapse, $N^a$ the shift, and $q_{ab}$ is the spatial metric induced on a three-dimensional hypersurface $\Sigma$.

The time-evolution vector field associated to this particular foliation is given by
\begin{equation}
    t^\mu = N n^\mu + N^a s_a^\mu
    \,,
    \label{eq:Time-evolution vector field}
\end{equation}
where $n^\mu$ is a unit vector normal to $\Sigma$, and $s^\mu_a$ are three basis vectors tangential to $\Sigma$.
The normal component of the time-evolution vector field is then given by the lapse, and its tangential components by the shift.
The index $\bar{0}$ will be used as the normal component of spacetime indices, and $t$ will be used for the time component.

We also define an internal normal vector $\hat{n}^I$ and internal spatial basis vectors $\hat{s}^I_i$, such that $\eta_{IJ}\hat{n}^I\hat{n}^J=-1$, $\eta_{IJ}\hat{s}^I_i\hat{s}^J_j=\delta_{ij}$, and $\eta_{IJ}\hat{n}^I\hat{s}^J_j=0$.
The Minkowski metric can be written in terms of these basis vectors as
\begin{equation}\label{eq:Internal foliation}
    \eta_{IJ} = - \hat{n}_I \hat{n}_J + \delta_{ij} \hat{s}^i_I \hat{s}^j_J\,,
\end{equation}
and the tetrad as \cite{Tetrads}
\begin{eqnarray}\label{eq:Tetrad coordinate basis}
    e^\mu_I
    \!\!&=&\!\! - \gamma N^{-1} \left(\hat{n}_I + v_i \hat{s}^i_I \right) t^\mu
    + \left(\gamma N^{-1} N^a + \varepsilon^a_k v^k\right) \hat{n}_I s^\mu_a
    \nonumber\\
    \!\!&&\!\!
    + \left(\gamma N^{-1} N^a v_i + \varepsilon^a_i \right) \hat{s}^i_I s^\mu_a
    \nonumber\\
    \!\!&=:&\!\! - n_I n^\mu + \varepsilon^a_i s^i_I s^\mu_a
    \,,
\end{eqnarray}
where $\gamma=1/\sqrt{1-v^2}$ and $v^2=v_iv^i$.
Notice that $e^t_0 = \hat{n}^I e^t_I = \gamma N^{-1}$ and $e^a_0 = \hat{n}^I e^a_I = - \left(\gamma N^{-1} N^a + \varepsilon^a_k v^k\right)$.
(We follow the index convention $T_0^I=\hat{n}^JT_J^I$ and $T^0_J = - \hat{n}_I T^I_J$.)
Inspection of the tetrad frame basis, defined by
\begin{equation}\label{eq:Tetrad frame}
    n_I = e^\mu_I n_\mu = \gamma \hat{n}_I + \gamma v_i \hat{s}^i_I\quad,\quad
    s^i_I = v^i \hat{n}_I + \hat{s}^i_I\,,
\end{equation}
reveals that it is boosted with respect to the internal one, defined by $(\hat{n}_I,\hat{s}^i_I)$, by the relative velocity $v^i$.

\subsection{Symplectic structure}
\label{sec:Symplectic structure}

Using $F^{KL}_{ta}\supset\dot{\omega}_a^{KL}$, the gravitational symplectic contribution to the action is given by
\begin{eqnarray}\label{eq:Symplectic structure - extended}
    \!\!&&\!\!
    \int {\rm d}^4x \left[\tilde{\cal P}^a_i \dot{K}_a^i + \tilde{\cal K}^a_i \dot{\Gamma}_a^i \right]
    \\
    \!\!&&\!\!\qquad\qquad\qquad
    = \int {\rm d}^4x \left[{\cal P}^a_i \dot{A}_a^i + {\cal K}^a_i \dot{B}_a^i \right]
    \nonumber\\
    \!\!&&\!\!\qquad\qquad\qquad
    = \int {\rm d}^4x \left[{\cal P}^a_i \dot{\cal D}_a^i + v_i \dot{\cal E}^i
    + \mathfrak{K}_{ij} \dot{\cal B}^{ij} \right]
    \,,\nonumber
\end{eqnarray}
where the configuration variables are the connection components
\begin{equation}\label{eq:Canonical connections}
    K^i_a =\omega^{0i}_a
    \quad,\quad
    \Gamma^i_a=\frac{1}{2} \tensor{\epsilon}{^i_k_l} \omega_a^{kl}\,,
\end{equation}
and their respective conjugate momenta are given by
\begin{eqnarray}\label{eq:Momenta Gauss}
    \tilde{\cal P}^a_i \!\!&=&\!\! {\cal P}^a_i+ \zeta {\cal K}^a_i
    \,,\\
    \tilde{\cal K}^a_i \!\!&=&\!\! {\cal K}^a_i - \zeta {\cal P}^a_i
    \,,\label{eq:Momenta Lorentz}
\end{eqnarray}
where
\begin{eqnarray}\label{eq:Momenta P}
    {\cal P}^a_i \!\!&=&\!\! \frac{|\det e|}{8\pi G} \left(e^t_0 e^a_i - e^t_i e^a_0\right)
    = \frac{|\det e|}{8\pi G} 2e^{[t}_0 e^{a]}_i
    \nonumber\\
    \!\!&=&\!\! {\rm sgn}(N) \frac{\gamma^2/(8\pi G)}{\det \varepsilon} \varepsilon^a_j \left(\delta^j_i - v^j v_i \right)
    \,,\\
    {\cal K}^a_i \!\!&=&\!\! \frac{|\det e|}{8\pi G} e^t_k e^a_l \tensor{\epsilon}{^k^l_i}
    \nonumber\\
    \!\!&=&\!\! {\rm sgn}(N) \frac{\gamma^2/(8\pi G)}{\det \varepsilon} \varepsilon^a_j \tensor{\epsilon}{_i^j^k} v_k
    \,.\label{eq:Momenta K}
\end{eqnarray}

The second and third lines of (\ref{eq:Symplectic structure - extended}) imply that there exist the following two canonical transformations, which turn out to be important in our canonical analysis.
The first one leads to the new configuration variables
\begin{eqnarray}\label{eq:Connection A}
    A_a^i \!\!&=&\!\! K^i_a
    - \zeta \Gamma^i_a\,,\\
    B_a^i \!\!&=&\!\! \Gamma^i_a+\zeta K^i_a
    \,,\label{eq:Connection B}
\end{eqnarray}
whose respective conjugate momenta are given by (\ref{eq:Momenta P}) and (\ref{eq:Momenta K}).
The second one leads to the new configuration variables
\begin{eqnarray}\label{eq:New connection D - extended}
    {\cal D}_a^i \!\!&=&\!\! A_a^i+ \tensor{\epsilon}{_j^i^k} v_k B_a^j
    + \mathfrak{K}^i_j B_a^j
    \,,\\
    {\cal E}^i \!\!&=&\!\! \tensor{\epsilon}{_j^k^i} B_a^j {\cal P}^a_k
    \,,\\
    {\cal B}_{ik} \!\!&=&\!\! B_b^j \delta_{j(i} {\cal P}^b_{k)}
    \,,
\end{eqnarray}
and the new momenta
\begin{eqnarray}\label{eq:K frak}
    v_i \!\!&=&\!\! - \frac{1}{2} \tensor{\epsilon}{_i_m^n} \left({\cal P}^{-1}\right)^m_b {\cal K}^b_n
    \,,\\
    \mathfrak{K}_{ij} \!\!&=&\!\! \left({\cal P}^{-1}\right)^m_b \delta_{m(i} {\cal K}^b_{j)}
    \,,
\end{eqnarray}
where $\left({\cal P}^{-1}\right)^i_a$ is the inverse of ${\cal P}^a_i$.

The fermionic contribution to the symplectic term is given by
\begin{eqnarray}\label{eq:Symplectic structure fermion}
    \int {\rm d}^4x \left[P_{\Psi{\dot{A}}} \dot{\Psi}^{\dot{A}} + \dot{\overline{\Psi}}_{\dot{A}} \overline{P_\Psi}{}^{\dot{A}} \right]
    \,,
\end{eqnarray}
with momenta
\begin{eqnarray}\label{eq:Momenta Dirac fermion}
    P_{\Psi} = - \overline{\Xi\Psi}
    \quad,\quad
    \overline{P_{\Psi}} = - \Xi \Psi
    \,,
\end{eqnarray}
where
\begin{eqnarray}\label{eq:Xi}
    \Xi \!\!&=&\!\! \frac{i}{2} |e| e^t_I \left(1-i\alpha\gamma^5\right) \gamma^I
    \\
    \!\!&=&\!\! \frac{i}{2} \gamma \sqrt{\det q} \left(1-i\alpha\gamma^5\right) \left(\gamma^0
    - v_i \gamma^i\right)
    \,.\nonumber
\end{eqnarray}

Being Grassmanian-valued, the fermionic variables obey the basic (anti-)Poisson brackets
\begin{eqnarray}
    \!\!\!\!\!
    \{\Psi^{\dot{A}}(x),P_{\Psi{\dot{B}}}(y)\} \!\!&=&\!\! \{P_{\Psi{\dot{B}}}(x),\Psi^{\dot{A}}(y)\} = \delta^{\dot{A}}_{\dot{B}} \delta^3(x-y)
    \\
    \!\!\!\!\!
    \{\overline{\Psi}_{\dot{A}}(x),\overline{P_{\Psi}}{}^{\dot{B}}(y)\} \!\!&=&\!\! \{\overline{P_{\Psi}}{}^{\dot{B}}(x),\overline{\Psi}_{\dot{A}}(y)\} = - \delta_{\dot{A}}^{\dot{B}} \delta^3(x-y)
    ,\nonumber
\end{eqnarray}
with all other combinations vanishing.

The full Poisson bracket between any two phase-space functionals ${\cal O}$ and ${\cal U}$ is given by
\begin{widetext}
\begin{eqnarray}\label{eq:Poisson bracket - extended phase space}
    \{{\cal O},{\cal U}\} \!\!&=&\!\! \int{\rm d}^3z \Bigg[\frac{\delta {\cal O}}{\delta A^k_c(z)} \frac{\delta {\cal U}}{\delta {\cal P}_k^c(z)}
    - \frac{\delta {\cal O}}{\delta {\cal P}^c_k(z)} \frac{\delta {\cal U}}{\delta A_c^k(z)}
    + \frac{\delta {\cal O}}{\delta B_c^k(z)} \frac{\delta {\cal U}}{\delta {\cal K}^c_k(z)}
    - \frac{\delta {\cal O}}{\delta {\cal K}^c_k(z)} \frac{\delta {\cal U}}{\delta B_c^k(z)}
    \\
    \!\!&&\!\!\qquad\qquad
    + \frac{\delta^R {\cal O}}{\delta \Psi^{\dot{A}}(z)} \frac{\delta {\cal U}}{\delta P_{\Psi{\dot{A}}}(z)}
    + \frac{\delta^R {\cal O}}{\delta P_{\Psi{\dot{A}}}(z)} \frac{\delta {\cal U}}{\delta \Psi^{\dot{A}}(z)}
    - \frac{\delta^R {\cal O}}{\delta \overline{\Psi}_{\dot{A}}(z)} \frac{\delta {\cal U}}{\delta \overline{P_\Psi}{}^{\dot{A}}(z)}
    - \frac{\delta^R {\cal O}}{\delta \overline{P_\Psi}{}^{\dot{A}}(z)} \frac{\delta {\cal U}}{\delta \overline{\Psi}_{\dot{A}}(z)}
    \Bigg]
    \nonumber
\end{eqnarray}
\end{widetext}
---where the superscript $R$ denotes a right derivative and left derivatives are always used otherwise---or a variation using other canonical phase-space coordinates.

The symplectic structure defined by the Poisson bracket (\ref{eq:Poisson bracket - extended phase space}) corresponds to an extended phase space whose number of canonical pairs exceeds that of the dynamical fields implied by the action.
Therefore, in order to describe the same theory of the Lagrangian formulation, a phase-space reduction must take place by imposing  a specific set of constraints that we identify in the next subsection.

\subsection{Non-dynamical variables and phase-space reduction}
\label{sec:Non-dynamical variables}

The 6 components $\omega_t^{IJ}$ are non-dynamical because their time derivatives do not appear in the action and imply 6 first-class constraints.
In the following, we will denote these components by
\begin{equation}
    K_t^i = \omega_t^{0i} \quad,\quad \Gamma_t^i = \frac{1}{2} \tensor{\epsilon}{^i_k_l} \omega^{kl}_t\,.
\end{equation}
Similarly, no time derivatives of the lapse and the shift appear in the action, leading to 4 additional first-class constraints.
The first-class status of these constraints is shown in the next section.

Furthermore, the relation
\begin{equation}\label{eq:KP relation}
    {\cal K}^a_i = \tensor{\epsilon}{_i^j^k} {\cal P}^a_j v_k\,,
\end{equation}
derived from comparing (\ref{eq:Momenta P}) and (\ref{eq:Momenta K}), implies that the momentum components $\mathfrak{K}_{ij}$ vanish on physical solutions,
\begin{equation}\label{eq:Second-class constraint K=0}
    \mathfrak{K}_{ij}=0\,,
\end{equation}
constituting 6 second-class constraints on the phase space.
This implies that the time derivatives of the components ${\cal B}_{ij}$ drop out from the symplectic term (\ref{eq:Symplectic structure - extended}), and hence constitute 6 additional non-dynamical components of the connection, which lead to another 6 second-class constraints
\begin{equation}\label{eq:Second-class constraint B=0}
    \frac{\delta {\cal H}}{\delta {\cal B}^{ij}} = 0\,,
\end{equation}
where ${\cal H}$ is the Hamiltonian.
This constitutes a secondary second-class constraint enforcing the preservation of (\ref{eq:Second-class constraint K=0}) under time evolution: $\delta {\cal H}/\delta {\cal B}^{ij}=-\{\mathfrak{K}_{ij},{\cal H}\}=0$.
The second-class status of these constraints and the fact that no tertiary constraints are necessary to enforce the preservation of (\ref{eq:Second-class constraint B=0}) under time evolution are  shown in Section~\ref{sec:Second-class constraints}.

The resulting 10 first-class constraints, together with the total of 12 second-class constraints, reduce the 18 gravitational canonical pairs to only 2 degrees of freedom on physical solutions, in agreement with our counting in the Lagrangian formulation.

On the other hand, the symplectic term (\ref{eq:Symplectic structure fermion}) implies that the configuration variables $\Psi$ and $\overline{\Psi}$ are kinematically independent fields, which in turn imply 16 canonical pairs of the fermion field.
However, imposing the reality conditions
\begin{equation}\label{eq:Reality conditions - canonical}
    \overline{\Psi}=\Psi^\dagger \beta
    \quad,\quad
    \overline{P_\Psi}=\beta P_\Psi^\dagger\,,
\end{equation}
lowers the total fermionic degrees of freedom from 16 to only 8 on dynamical solutions---the reality conditions can be consistently imposed if all the constraints are Hermitian, a property that will be applied in the forthcoming canonical decomposition and hence will always be assumed to hold.
If the Hamiltonian is a fermion bilinear with up to first-order derivatives of the fermionic variables---which is the case in our system as shown in the next section---then Hamilton's equations of motion for $\Psi$ and $P_\Psi$ result in first-order differential equations of $\Psi$ and $P_\Psi$, 
\begin{equation}
    O \Psi = 0
\end{equation}
and
\begin{equation}
    P_\Psi U = 0\,,
\end{equation}
with some linear operators $O$ and $U$, respectively.
This implies that the 8 components of $\Psi$ reduce to 4 degrees of freedom but so will the components of $P_\Psi$, totaling 8 fermionic degrees of freedom, rather than the 4 implied by the action.
The reason for the doubling of the fermionic degrees of freedom can be identified as the fact that the momentum $P_\Psi$ is not related to $\overline{\Psi}$ by the relation (\ref{eq:Momenta Dirac fermion})---and their conjugate versions---at the kinematical level.
These relations can hold on dynamical solutions, but only as a special case where they are restricted to satisfy the relation (\ref{eq:Momenta Dirac fermion}) and its conjugated version because they do not arise from the solution to the linear equations of motion alone: This requires the imposition of 8 additional constraints, given by
\begin{equation}\label{eq:Dirac fermion constraint - 1}
    C_{\overline{\Psi}} = P_\Psi + \overline{\Psi}\,\overline{\Xi}
\end{equation}
and
\begin{equation}\label{eq:Dirac fermion constraint - 2}
    C_\Psi = \overline{P_\Psi} + \Xi \Psi
    \,,
\end{equation}
which, depending on the details of the first-class constraints, may be second class.
We reiterate that the imposition of these constraints is different from that of the reality conditions (\ref{eq:Reality conditions - canonical}) and must be implemented in addition to the latter to describe Dirac fermions.

If the constraints (\ref{eq:Dirac fermion constraint - 1}) and (\ref{eq:Dirac fermion constraint - 2}) are not imposed, then the canonical system implies a more general theory with 4 additional fermionic degrees of freedom compared to the Dirac theory that still satisfies the reality conditions.
Such a theory is instead based on a different action contribution, given by
\begin{eqnarray}\label{eq:Extended Dirac action}
    S_F \!\!&=&\!\! \int {\rm d}^4x |e| \frac{i}{2} e^\mu_I \Big[\overline{\Phi} \gamma^I \left(1-i\alpha\gamma^5\right) D_\mu \Psi
    \\
    \!\!&&\!\!\qquad\qquad\qquad\qquad
    - D_\mu \overline{\Psi} \left(1-i\alpha\gamma^5\right) \gamma^I \Phi\Big]
    \,,\nonumber
\end{eqnarray}
where $\Phi$ is an independent fermion field.
The fermionic equations of motion of this action, $\delta S_F/\delta \overline{\Phi}=0$ and $\delta S_F/\delta \overline{\Psi}=0$, are respectively given by
\begin{equation}\label{eq:Extended fermion EoM Psi}
    |e| \frac{i}{2} e^\mu_I \gamma^I \left(1-i\alpha\gamma^5\right) D_\mu \Psi = 0
\end{equation}
and
\begin{eqnarray}\label{eq:Extended fermion EoM Phi}
    \!\!&&\!\!
    |e| \frac{i}{2} e^\mu_I \left(1-i\alpha\gamma^5\right) \gamma^I D_\mu \Phi
    \\
    \!\!&&\!\!\quad
    + \frac{i}{2} D_\mu \left(|e| e^\mu_I\right) \left(1-i\alpha\gamma^5\right) \gamma^I \Phi
    = 0
    \,,\nonumber
\end{eqnarray}
which differ from (\ref{eq:Dirac equation}) and the solutions are independent of $\alpha$, up to its possible appearance in $D_\mu \left(|e| e^\mu_I\right)$.
The covariant divergence of the tetrad based on this extended action is richer than that of the Dirac action:
\begin{widetext}
\begin{eqnarray}\label{eq:Divergence densitized tetrad - extended}
    D_\mu \left(|e|e^\mu_I\right) \!\!&=&\!\! \frac{8\pi G}{4} \frac{|e|}{1+\zeta^2} \frac{1}{2} \left(\frac{\zeta}{2} \tensor{\epsilon}{_I_N^R^S} \overline{\Phi} \left(1+i\alpha\gamma^5\right) [\gamma^N,S_{RS}]_+ \Psi
    + \overline{\Phi} \left(1+i\alpha\gamma^5\right) [\gamma^N,S_{IN}] \Psi
    + c.c.\right)
    \nonumber\\
    \!\!&=&\!\! - \frac{3}{8} 8\pi G \frac{|e|}{1+\zeta^2} \left(\overline{\Phi} \gamma_I \left[\left(\zeta-\alpha\right) \gamma^5
    - i \left(1+\alpha\zeta\right)\right]\Psi
    + c.c.\right)
    \,.
\end{eqnarray}
\end{widetext}

The symplectic term of the action contribution (\ref{eq:Extended Dirac action}) is given by (\ref{eq:Symplectic structure fermion}) too, but with the momentum
\begin{equation}\label{eq:Spinor momentum Phi}
    P_{\Psi} = - \overline{\Xi\Phi} \,.
\end{equation}
Notice, however, that, imposing $\Phi=\Psi$ a posteriori, equations (\ref{eq:Extended fermion EoM Psi}) and (\ref{eq:Extended fermion EoM Phi}) would imply that $D_\mu \left(|e| e^\mu_I\right)$ vanishes; however, according to (\ref{eq:Divergence densitized tetrad - extended}), $D_\mu \left(|e| e^\mu_I\right)$ is generally non-vanishing with the exceptions lying on a specific small set of configurations of the fermionic fields.
Solutions admitting $\Phi=\Psi$ are therefore extremely limited.
Hence, the implementation of the Dirac-fermion constraints (\ref{eq:Dirac fermion constraint - 1}) and (\ref{eq:Dirac fermion constraint - 2}) must be performed a priori as a phase-space reduction rather than as a restriction to the solutions to the equations of motion of the extended phase space to recover a Dirac-fermion system---this is analogous to a free two-dimensional single-particle system with its motion confined to a ring: Circular motion cannot be recovered by restricting the solutions (linear motion) of the free single-particle system; the phase-space reduction must be performed prior to deriving the equations of motion from the Lagrangian or, alternatively, by using Lagrange multipliers to constrain the motion.
This can be achieved in the canonical formulation by use of Dirac brackets, following the treatment of second-class constraints.

We conclude that the 34 canonical pairs of the extended phase space described by (\ref{eq:Poisson bracket - extended phase space}) reduce to only 6 degrees of freedom (2 of which are gravitational and 4 are fermionic) upon the imposition of the reality conditions (\ref{eq:Reality conditions - canonical}), the Dirac-fermion constraints (\ref{eq:Dirac fermion constraint - 1}) and (\ref{eq:Dirac fermion constraint - 2}), and the gravitational second-class constraints (\ref{eq:Second-class constraint K=0}) and (\ref{eq:Second-class constraint B=0}), as well as the first-class constraints (and use of the equations of motion they generate) discussed in the next section.
The implementation of the second-class constraints via Dirac brackets is discussed in Section~\ref{sec:Second-class constraints}.

\subsection{Canonical decomposition of geometric quantities}

Having identified the canonical phase-space variables in terms of the Lagrangian fields, it is useful to invert these relations for a systematic canonical decomposition of the action.
We gather several of these relations in this subsection.

The inverse spatial metric is given by
\begin{eqnarray}\label{eq:Inverse spatial metric - canonical}
    q^{ab}
    \!\!&=&\!\! \frac{|\det {\cal P}|^{-1}}{8\pi G} \left(\delta^{ij}+\gamma^2 v^iv^j \right) {\cal P}^a_i {\cal P}^b_j
    \\
    \!\!&=&\!\!
    \frac{\gamma^2 |\det {\cal P}|^{-1}}{8\pi G} \delta^{ij} \left({\cal P}^a_i {\cal P}^b_j-{\cal K}^a_i {\cal K}^b_j\right) \bigg|_{\mathfrak{K}=0}
    \,.\nonumber
\end{eqnarray}
The determinant of $q_{ab}$ is then given by
\begin{equation}
    \det q
    = (8\pi G)^3 \frac{|\det {\cal P}|}{\gamma^2}
    \,.
\end{equation}

In terms of the new canonical variables, the covariant derivative of the fermion field is given by
\begin{eqnarray}
    D_t \Psi \!\!&=&\!\! \dot{\Psi} - i K_t^i S_{0i} \Psi - i \Gamma_t^i S_i \Psi
    \,,\\
    D_a \Psi \!\!&=&\!\! \partial_a \Psi - i K^i_a S_{0i} \Psi - i \Gamma^i_a S_i \Psi
    \,,
\end{eqnarray}
where we defined
\begin{equation}
    S^i=\frac{1}{2}\tensor{\epsilon}{^i_j_k}S^{jk}\,.
\end{equation}

The identity
\begin{equation}\label{eq:Gamma boost relation}
    \left(\gamma^0
    - v_i \gamma^i\right) \left(\gamma^0
    - v_j \gamma^j\right) = \frac{1}{\gamma^2}
\end{equation}
is useful to invert (\ref{eq:Momenta Dirac fermion}) into
\begin{eqnarray}
    \overline{\Psi}
    \!\!&=&\!\! - P_\Psi \overline{\Xi^{-1}}
    \,,\\
    \Psi \!\!&=&\!\! - \Xi^{-1} \overline{P_\Psi}
    \,,
\end{eqnarray}
where
\begin{equation}
    \Xi^{-1} = - \frac{2 i}{1+\alpha^2} \frac{\gamma}{\sqrt{\det q}} \left(\gamma^0
    - v_i \gamma^i\right) \left(1+i\alpha\gamma^5\right)
\end{equation}
is the inverse of $\Xi$.

It is useful to write the Dirac equation (\ref{eq:Dirac equation}) in terms of the canonical variables because it serves as a guide to construct the Hamiltonian that can generate it:
\begin{eqnarray}\label{eq:Dirac equation - Lagrange - phase space}
    \dot{\Psi}
    \!\!&=&\!\!
    \Theta^a \partial_a \Psi
    + i \Omega \Psi
    \\
    \!\!&&\!\!
    - \frac{\gamma^2\left(\gamma^0-v_i\gamma^i\right) \left(1-i\alpha\gamma^5\right) D_\mu \left(|e| e^\mu_I \gamma^I\right) \Psi}{2(8\pi G)^{3/2} \sqrt{|\det {\cal P}|}}
    \,,\nonumber
\end{eqnarray}
where\begin{equation}
    \Theta^a = N^a
    + \frac{8\pi G N}{\sqrt{\det q}} {\cal P}^a_q \left[v^q
    - \left(\gamma^0
    - v_k \gamma^k\right) \gamma^q\right]
    \,,
\end{equation}
\begin{eqnarray}
    \!\!
    \Omega \!\!&=&\!\! - \left[K_t^i - \left(N^a+N \frac{8\pi G}{\sqrt{\det q}} v^q {\cal P}^a_q\right) K_a^i\right] S_{0i}
    \\
    \!\!&&\!\!
    - \left[\Gamma_t^i - \left(N^a+N \frac{8\pi G}{\sqrt{\det q}} v^q {\cal P}^a_q\right) \Gamma_a^i\right] S_i
    \nonumber\\
    \!\!&&\!\!
    - N \frac{8\pi G}{\sqrt{\det q}} \left(\gamma^0-v_m\gamma^m\right) {\cal P}^a_j \gamma^j \left(K_a^i S_{0i}
    + \Gamma_a^i S_i\right)
    \,,\nonumber
\end{eqnarray}
and
\begin{eqnarray}\label{eq:Covariant divergence tetrad - phase space}
    D_\mu \left(|e| e^\mu_I \gamma^I\right)
    \!\!&=&\!\!
    \frac{2}{i} \frac{1+i\alpha\gamma^5}{1+\alpha^2} \Big[\{\Xi,{\cal H}\}
    - \partial_a \left(\overline{\Theta^a} \Xi\right)
    \nonumber\\
    \!\!&&\!\!\qquad\qquad\quad
    - i \left(\overline{\Omega}\Xi-\Xi \Omega\right)\Big]
    \,,
\end{eqnarray}
where
\begin{widetext}
\begin{eqnarray}
    \overline{\Omega}\Xi-\Xi \Omega
    \!\!&=&\!\!
    \frac{i}{2} \gamma \sqrt{\det q} \left(1+i\alpha\gamma^5\right) \Bigg[
    - i \left[K_t^i - \left(N^a+N \frac{8\pi G}{\sqrt{\det q}} v^q {\cal P}^a_q\right) K_a^i\right] \left(\gamma_i
    - v_i \gamma^0\right)
    \\
    \!\!&&\!\!\qquad\qquad\qquad
    + i \left[\Gamma_t^i - \left(N^a+N \frac{8\pi G}{\sqrt{\det q}} v^q {\cal P}^a_q\right) \Gamma_a^i\right] v_m \tensor{\epsilon}{_i^m_l} \gamma^{l}
    - i \frac{8\pi G N}{\gamma^2\sqrt{\det q}} {\cal P}^a_j \left(K_a^j \gamma^0
    + \Gamma_a^i \tensor{\epsilon}{_i^j_l} \gamma^{l}\right)
    \Bigg]
    \,.\nonumber
\end{eqnarray}
\end{widetext}
The bracket $\{\Xi,{\cal H}\}$ cannot be written definitively in terms of the phase-space variables at this stage because it requires knowledge of the Hamiltonian, yet to be computed.
However, on shell, it must be such that Eq.~(\ref{eq:Covariant divergence tetrad - phase space}) coincides with (\ref{eq:Divergence densitized tetrad}) if the Dirac-fermion constraints are imposed.

Using the above results, the equation of motion for the fermionic momentum is given by
\begin{eqnarray}\label{eq:Dirac equation momentum - phase space}
    \dot{P}_\Psi
    \!\!&=&\!\!
    \partial_a \left(P_\Psi \Theta^a\right)
    - i P_\Psi \Omega
    \\
    \!\!&&\!\!
    + \frac{\gamma P_\Psi \left(1-i\alpha\gamma^5\right) \left(\gamma^0
    - v_n \gamma^n\right) D_\mu \left(|e| e^\mu_I \gamma^I\right)}{\sqrt{\det q}}
    \,.\nonumber
\end{eqnarray}

Finally, relevant components of the torsion equation (\ref{eq:Torsion EoM}) for the forthcoming canonical analysis can be expressed by the symmetric internal tensor
\begin{eqnarray}\label{eq:Torsion-spatial-sym}
    {\cal T}^{ij}
    \!\!&\equiv&\!\! - \frac{\delta_k^{(i} \tensor{\epsilon}{^{j)}^p^q} {\cal P}^a_p {\cal P}^b_q}{\sqrt{8\pi G |\det {\cal P}|}} \tensor{{\cal T}}{^k_a_b}
    \\
    \!\!&=&\!\!
    \left(\delta^{ij}\delta^p_q-\delta^{p(i}\delta_q^{j)}\right) {\cal P}^d_p \Gamma_d^q
    - \delta_q^{(i}\delta^{j)p} {\cal K}^d_p K_d^q
    \nonumber\\
    \!\!&&\!\!
    + \delta_k^{(i} \tensor{\epsilon}{^{j)}^p^q} {\cal P}^c_p {\cal P}^d_q \partial_d \left({\cal P}^{-1}\right)_c^k
    \nonumber\\
    \!\!&&\!\!
    + \frac{1}{2} \frac{1}{1+\zeta^2} \left[P_\Psi \left(\zeta - i \gamma^5\right) \upsilon^{ij} \Psi
    + c.c.
    \right]
    \,,\nonumber
\end{eqnarray}
where
\begin{equation}
    \upsilon^{ij}=\delta^{ij}+\gamma^2 \left(\gamma^0 - v_u \gamma^u\right) v^{(i} \gamma^{j)}\,.
\end{equation}

\section{First-class constraints}
\label{sec:First-class constraints}

\subsection{Constraints}

The action $S=S_G+S_F$ can be written as
\begin{eqnarray}\label{eq:Action canonical 1}
    S
    \!\!\!&=&\!\!\! \int {\rm d}^4x \bigg[\tilde{\cal P}^a_i F_{ta}^{0i}
    + \tilde{\cal K}^a_i {\cal F}_{ta}^i
    + P_\Psi D_t \Psi
    + D_t \overline{\Psi}\,\overline{P_\Psi}
    \nonumber\\
    \!\!\!&&\!\!\!\qquad\quad
    + \frac{|e|}{16\pi G} e^a_I e^b_J \tensor{P}{^I^J_K_L} F^{KL}_{ab}
    \nonumber\\
    \!\!\!&&\!\!\!\qquad\quad
    + \frac{i}{2} |e| e^a_I \left(\overline{\Psi} \gamma^I D_a \Psi
    - D_a \overline{\Psi} \gamma^I \Psi\right) \bigg]
    \nonumber\\
    \!\!\!&=&\!\!\! \int {\rm d}^4x \left[
    \tilde{\cal P}^a_i \dot{K}_a^i + \tilde{\cal K}^a_i \dot{\Gamma}_a^i 
    + P_\Psi \dot{\Psi}
    + \dot{\overline{\Psi}}\,\overline{P_\Psi} \right]
    \nonumber\\
    \!\!\!&&\!\!\!
    - H[|N|]
    - H_a[N^a]
    - L_i[K_t^i]
    - G_i[\Gamma_t^i]
    \,,
\end{eqnarray}
which is linear in all Lagrange multipliers and, therefore, all the local expressions of $H$, $H_a$, $L_i$, and $G_i$ are constraints.
Only the first line contributes to the symplectic term and to $L_i$ and $G_i$, while only the second and third lines contribute to $H$ and $H_a$.
Notice that the absolute value of the lapse $|N|$ smears the Hamiltonian constraint; this is because the action depends on the absolute value of the co-tetrad's determinant $|e|$.
In the following, however, we will often write $H[N]$, but it is understood that we use only its magnitude.
This detail is not relevant in most applications because the gauge choices of the lapse that change sign would require that it vanishes in some region of the spacetime, which would amount to a limitation of the coordinate chart related to that specific gauge choice and hence another well-defined gauge, with locally non-vanishing lapse, is required to describe that region of the spacetime; however, it will be important for the discussion on discrete symmetries presented in Section~\ref{sec:Discrete symmetries}.

We first focus on the Lorentz--Gauss constraints, given by $L_i=L_i^{(G)}+L_i^{(F)}$ and $G_i=G_i^{(G)}+G_i^{(F)}$, respectively, where
\begin{eqnarray}\label{eq:Lorentz constraint - grav}
    L_i^{(G)} \!\!&=&\!\! - \partial_a \tilde{\cal P}^a_i
    - \left( \tilde{\cal P}^a_k \Gamma^l_a
    - \tilde{\cal K}^a_k K^l_a \right)\tensor{\epsilon}{^k_l_i}
    \\
    \!\!&=&\!\! - \partial_a {\cal P}^a_i
    - \zeta \partial_a {\cal K}^a_i
    - \left({\cal P}^a_k B_a^l
    - {\cal K}^a_k A_a^l\right)\tensor{\epsilon}{^k_l_i}
    \,,\nonumber
\end{eqnarray}
\begin{equation}\label{eq:Lorentz constraint - fermion}
    L_i^{(F)} =
    - i \left[P_\Psi S_{0i} {\Psi} - \overline{\Psi}\, S_{0i} \overline{P_\Psi} \right]
    \,,
\end{equation}
\begin{eqnarray}\label{eq:Gauss constraint - grav}
    G_i^{(G)} \!\!&=&\!\! - \partial_a \tilde{\cal K}^a_i
    - \left(\tilde{\cal P}^a_k K^l_a
    + \tilde{\cal K}^a_k \Gamma^l_a \right) \tensor{\epsilon}{^k_l_i}
    \\
    \!\!&=&\!\! - \partial_a {\cal K}^a_i
    + \zeta \partial_a {\cal P}^a_i
    - \left({\cal P}^a_k A_a^l
    + {\cal K}^a_k B_a^l \right) \tensor{\epsilon}{^k_l_i}
    \,,\nonumber
\end{eqnarray}
and
\begin{equation}\label{eq:Gauss constraint - fermion}
    G_i^{(F)} = - i \left[P_\Psi S_i {\Psi} - \overline{\Psi}\,S_i \overline{P_\Psi} \right]
    \,.
\end{equation}

The gravitational contributions to these constraints generate proper Lorentz transformations of the triad $e^a_I$ and the spatial strength tensor $F^{IJ}_{ab}$ even off shell \cite{Tetrads}.
The fermionic contributions preserve the generation of Lorentz transformations of the gravitational variables and also generate those of the fermionic variables:
\begin{equation}
    \{\Psi,L_i[\beta^i]\} = - i \beta^i S_{0i} {\Psi}
    \quad,\quad
    \{\Psi,G_i[\theta^i]\} = - i \theta^i S_i {\Psi}\,,
\end{equation}
\begin{equation}
    \{P_\Psi,L_i[\beta^i]\} = i \beta^i P_\Psi S_{0i}
    \quad,\quad
    \{P_\Psi,G_i[\theta^i]\} = i \theta^i P_\Psi S_i \,.
\end{equation}

The vector constraint is given by $H_a=H^{(G)}_a+H^{(F)}_a$, where
\begin{eqnarray}\label{eq:Vector constraint}
    H_a^{(G)}
    \!\!&=&\!\! \tilde{\cal P}^b_i F^{0i}_{ab}
    + \tilde{\cal K}^b_i {\cal F}^i_{ab}
    \\
    \!\!&=&\!\! {\cal P}^b_i \left(F^{0i}_{ab} - \zeta {\cal F}^i_{ab}\right)
    + {\cal K}^b_i \left({\cal F}^i_{ab} + \zeta F^{0i}_{ab}\right)
    \,,\nonumber
\end{eqnarray}
is the gravitational contribution and
\begin{eqnarray}
    H_a^{(F)} \!\!&=&\!\! P_\Psi D_a \Psi
    + \overline{D_a \Psi}\; \overline{P_\Psi}
    \\
    \!\!&=&\!\! P_\Psi \partial_a \Psi
    + \partial_a \overline{\Psi}\; \overline{P_\Psi}
    - K_a^i L^{(F)}_i
    - \Gamma_a^i G^{(F)}_i
    \,,\nonumber
\end{eqnarray}
is the fermionic contribution.

Above, we used the expressions
\begin{equation}
    F_{ab}^{0i} = 2 \partial_{[a} K_{b]}^i - 2 \tensor{\epsilon}{^i_j_k} K_{[a}^j \Gamma_{b]}^{k}
\end{equation}
and
\begin{eqnarray}
    {\cal F}^i_{ab}
    \!\!&=&\!\! \frac{1}{2} \tensor{\epsilon}{^i_k_l} F^{kl}_{ab}
    \nonumber\\
    \!\!&=&\!\! 2 \partial_{[a} \Gamma_{b]}^i + \tensor{\epsilon}{^i_j_k} K_a^j K_b^k - \tensor{\epsilon}{^i_j_k} \Gamma_a^j \Gamma_b^k\,.
\end{eqnarray}
The linear combinations with the Barbero--Immirzi parameter can be written more compactly in the $A,B$ variables,
\begin{eqnarray}\label{eq:F lin comb - 1}
    \!\!\!\!&&\!\!\!\!
    F^{0i}_{ab} - \zeta {\cal F}^i_{ab}
    =
    2 \partial_{[a} A_{b]}^i
    \\
    \!\!\!\!&&\!\!\!\!\qquad
    + 2 \tensor{\epsilon}{^i_k_l} \frac{1}{1+\zeta^2} \left( \frac{\zeta}{2} \left(A_{[a}^k A_{b]}^l - B_{[a}^k B_{b]}^l\right)
    - A_{[a}^k B_{b]}^l \right)
    \,,\nonumber
\end{eqnarray}
\begin{eqnarray}
    \!\!\!\!&&\!\!\!\!
    \label{eq:F lin comb - 2}
    {\cal F}^i_{ab} + \zeta F^{0i}_{ab}
    = 2 \partial_{[a} B_{b]}^i
    \\
    \!\!\!\!&&\!\!\!\!\qquad
    + 2 \tensor{\epsilon}{^i_k_l} \frac{1}{1+\zeta^2} \left( \frac{1}{2} \left(A_{[a}^k A_{b]}^l - B_{[a}^k B_{b]}^l\right)
    + \zeta A_{[a}^k B_{b]}^l \right)
    \,.\nonumber
\end{eqnarray}

The Hamiltonian constraint is given by
\begin{equation}\label{eq:Hamiltonian constraint - grav}
    H = H^{(1)} + H^{(2)}
    \,,
\end{equation}
where
\begin{equation}\label{eq:H1}
    H^{(1)} = H_a \frac{\gamma \sqrt{(\det {\cal P})^{-1}}}{\sqrt{8\pi G}} v^q {\cal P}^a_q
    \,.\nonumber
\end{equation}
and $H^{(2)}=H^{(2)}_{(G)}+H^{(2)}_{(F)}$ with the gravitational contribution
\begin{eqnarray}\label{eq:H2 - grav}
    H^{(2)}_{(G)} \!\!&=&\!\! - \frac{\sqrt{|(\det {\cal P})^{-1}|}}{2\gamma\sqrt{8\pi G}} {\cal P}^a_p {\cal P}^b_q \tensor{\epsilon}{^p^q_i} \left({\cal F}^i_{ab} + \zeta F^{0i}_{ab}\right)
    \nonumber\\
    \!\!&&\!\!
    + \frac{\sqrt{|\det {\cal P}}|}{\gamma} \sqrt{8\pi G}\Lambda
    \,.
\end{eqnarray}
and the fermionic contribution
\begin{widetext}
\begin{eqnarray}\label{eq:HF2}
    H_{(F)}^{(2)}\!\!&=&\!\!
    - \frac{\gamma \sqrt{(\det {\cal P})^{-1}}}{\sqrt{8\pi G}} {\cal P}^a_j \Big[P_\Psi \left(\gamma^0-v_m\gamma^m\right) \gamma^j \left(\partial_a \Psi + i \left(K_a^i S_{0i} + \Gamma_a^i S_i\right) \Psi\right)
    \nonumber\\
    \!\!&&\!\!\qquad\qquad\qquad\qquad\quad
    + \left(\partial_a \overline{\Psi} - i \overline{\Psi} \left(K_a^i S_{0i} + \Gamma_a^i S_i\right) \right) \gamma^j \left(\gamma^0-v_m\gamma^m\right) \overline{P_\Psi} \Big]\,.
\end{eqnarray}
\end{widetext}
(The various replacements of $\overline{\Psi}$ for $P_\Psi$, and their respective conjugates, are necessary to recover the correct equations of motion and to respect the discrete symmetries of the theory as will be clear in Section~\ref{sec:Discrete symmetries}.)

Because all of the first-class constraints are independent of $\alpha$, we find that the theory is kinematically independent of the non-minimal coupling parameter.

\subsection{Constraint algebra}
\label{sec:Constraint algebra}

A straightforward computation shows that $L_i$ and $G_i$ satisfy the Lorentz algebra,
\begin{eqnarray}
    \{L_i[K_t^i],L_j[\beta^j]\} \!\!&=&\!\! - G_k [\tensor{\epsilon}{^k_i_j} K_t^i \beta^j]
    \,,\\
    \{G_i[\Gamma_t^i],L_j[\beta^j]\} \!\!&=&\!\! L_k[\tensor{\epsilon}{^k_i_j} \Gamma_t^i \beta^j]
    \,,\\
    \{G_i[\Gamma_t^i],G_j[\theta^j]\} \!\!&=&\!\! G_k [ \tensor{\epsilon}{^k_i_j}\Gamma_t^i \theta^j]\,.
\end{eqnarray}

Now consider the linear combination $\mathfrak{D}_b[N^b]=H_a[N^b]+G_i[\Gamma_b^iN^b]+L_i[K_b^iN^b]$; in its local form, it is given by
\begin{eqnarray}
\label{eq:Spatial diffeomorphism generator}
    \mathfrak{D}_a
    \!\!&=&\!\! {\cal P}^b_i \partial_a A_b^i
    + {\cal K}^b_i \partial_a B_b^i
    - \partial_b \left({\cal P}^b_i A_a^i+{\cal K}^b_i B_a^i\right)
    \nonumber\\
    \!\!&&\!\!
    + P_\Psi \partial_a \Psi
    + \partial_a \overline{\Psi}\, \overline{P_\Psi}
    \,.
\end{eqnarray}

It is then clear that the functional $\mathfrak{D}_a[N^a]$ generates spatial diffeomorphisms of all phase-space variables.
It follows that
\begin{eqnarray}\label{eq:Vector constraint bracket}
    \{{\cal O},H_b[N^b]\} 
    \!\!&=&\!\! \mathcal{L}_{\vec N} {\cal O}
    + \{G_i\left[\Gamma_b^iN^b\right],{\cal O}\}
    \nonumber\\
    \!\!&&\!\!
    + \{L_i\left[K_b^iN^b\right],{\cal O}\}\,,
\end{eqnarray}
for any phase-space function ${\cal O}$.

Using the above, we obtain that the vector constraint commutes with the Lorentz--Gauss constraints,
\begin{eqnarray}\label{eq:HaJ}
    \{H_a [N^a],G_k[\theta^k]\} \!\!&=&\!\! 0\,,\\
    \{H_a [N^a],L_k[\beta^k]\} \!\!&=&\!\! 0\,.\label{eq:HaL}
\end{eqnarray}
Similarly, using (\ref{eq:Vector constraint bracket}), together with (\ref{eq:HaJ}) and (\ref{eq:HaL}), we obtain
\begin{eqnarray}\label{eq:HaHa}
    \{H_a[N^a],H_c[\epsilon^c]\}
    \!\!&=&\!\! - H_a \left[\mathcal{L}_{\vec \epsilon} N^a\right]
    - G_i\left[\epsilon^a N^b {\cal F}^i_{ab}\right]
    \nonumber\\
    \!\!&&\!\!
    - L_i\left[\epsilon^a N^b F^{0i}_{ab}\right]\,.
\end{eqnarray}

The second and third lines of the action (\ref{eq:Action canonical 1}) are Lorentz invariant and they equal $-H[N]-H_a[N^a]$.
Using (\ref{eq:HaJ}) and (\ref{eq:HaL}), as well as the fact that the Lorentz--Gauss constraints generate proper Lorentz transformations on and off shell, it follows that the Lorentz--Gauss constraints commute with the Hamiltonian constraint too:
\begin{eqnarray}\label{eq:HJ}
    \{H [N],G_k[\theta^k]\} \!\!&=&\!\! 0\,,\\
    \{H [N],L_k[\beta^k]\} \!\!&=&\!\! 0\,.\label{eq:HL}
\end{eqnarray}

Using (\ref{eq:Vector constraint bracket}), we obtain the bracket
\begin{eqnarray}
    \{H[N],H_c[\epsilon^c]\}
    \!\!&=&\!\! - H[\mathcal{L}_{\vec \epsilon} N]
    + G_i\left[N\epsilon^a {\cal F}^i_{\bar{0}a}\right]
    \nonumber\\
    \!\!&&\!\!
    + L_i\left[N\epsilon^a F^{0i}_{\bar{0}a}\right]\,,
\end{eqnarray}
where we defined the expressions
\begin{eqnarray}
    F_{\bar{0}a}^{0i} \!\!&=&\!\! \frac{1}{N} \{K_a^i,H[N]\}
    \nonumber\,,\\
    {\cal F}_{\bar{0}a}^i \!\!&=&\!\! \frac{1}{N} \{\Gamma_a^i,H[N]\}\,,\label{eq:calFi}
\end{eqnarray}
which are compatible with the identification of the normal components of the strength tensor,
\begin{eqnarray}
    n^\mu F^{0i}_{\mu a} \!\!&=&\!\! \frac{1}{N} F_{ta}^{0i} - \frac{N^b}{N} F_{ba}^{0i}
    =
    \frac{1}{N} \{K_a^i,H[N]\}
    \nonumber\,,\\
    n^\mu {\cal F}_{\mu a}^i \!\!&=&\!\! \frac{1}{N} {\cal F}_{ta}^i - \frac{N^b}{N} {\cal F}_{ba}^i
    = \frac{1}{N} \{\Gamma_a^i,H[N]\}\,.
\end{eqnarray}

Using all the above, we compute the bracket of the Hamiltonian constraint with itself in steps.
First, using (\ref{eq:HaHa}) and (\ref{eq:Vector constraint bracket}), we obtain
\begin{equation}
    \{H^{(1)}[N],H^{(1)}[\epsilon^{\bar{0}}]\}
    = 0
    \,,
\end{equation}
and
\begin{eqnarray}
    \!\!&&\!\!\{H^{(2)}[N],H^{(1)}[\epsilon^{\bar{0}}]\}+\{H^{(1)}[N],H^{(2)}[\epsilon^{\bar{0}}]\}
    \\
    \!\!&&\!\!
    \qquad= - H^{(2)} \left[\frac{\sqrt{(\det {\cal P})^{-1}}}{\sqrt{8\pi G}} \gamma v^q {\cal P}^a_q (\epsilon^{\bar{0}}\partial_a N-N\partial_a \epsilon^{\bar{0}})\right]
    \nonumber\\
    \!\!&&\!\!
    \qquad\quad
    - H_a\left[q^{ab} \left(\epsilon^{\bar{0}}\partial_bN-N\partial_b\epsilon^{\bar{0}}\right)\right]
    \,,\nonumber
\end{eqnarray}
where $q^{ab}$ is the inverse spatial metric (\ref{eq:Inverse spatial metric - canonical}).
Using the antisymmetry of the Poisson brackets---with proper care of the sign changes from commuting the fermionic variables---it follows that
\begin{eqnarray}
    \!\!&&\!\!\{H^{(2)}_{(G)}[N],H^{(2)}_{(G)}[\epsilon^{\bar{0}}]\}
    \\
    \!\!&&\!\!
    \qquad=
    H^{(2)}_{(G)}\left[\frac{\sqrt{(\det {\cal P})^{-1}}}{\sqrt{8\pi G}} \gamma v^q {\cal P}^b_q \left(\epsilon^{\bar{0}}\partial_bN-N\partial_b\epsilon^{\bar{0}}\right)\right]\,,\nonumber
\end{eqnarray}
\begin{equation}
    \{H^{(2)}_{(F)}[N],H^{(2)}_{(F)}[\epsilon^{\bar{0}}]\} = 0\,,
\end{equation}
and
\begin{eqnarray}
    \!\!&&\!\!
    \{H^{(2)}_{(G)}[N],H^{(2)}_{(F)}[\epsilon^{\bar{0}}]\}
    + \{H^{(2)}_{(F)}[N],H^{(2)}_{(G)}[\epsilon^{\bar{0}}]\}
    \\
    \!\!&&\!\!\qquad
    = H^{(2)}_{(F)}\left[\frac{\sqrt{(\det {\cal P})^{-1}}}{\sqrt{8\pi G}} \gamma v^q {\cal P}^b_q \left(\epsilon^{\bar{0}}\partial_bN-N\partial_b\epsilon^{\bar{0}}\right)\right]
    \,.\nonumber
\end{eqnarray}
Therefore, the full bracket of the Hamiltonian constraint with itself is given by
\begin{equation}\label{eq:HH}
    \{H[N],H[\epsilon^{\bar{0}}]\}=- H_a\left[q^{ab} \left(\epsilon^{\bar{0}}\partial_bN-N\partial_b\epsilon^{\bar{0}}\right)\right]\,.
\end{equation}

In summary, the constraint algebra in the extended phase space is given by
\begin{eqnarray}
    \label{eq:HaHa-Poisson}
    \{H_a[N^a],H_c[\epsilon^c]\}
    \!\!&=&\!\! - H_a \left[\mathcal{L}_{\vec \epsilon} N^a\right]
    - G_i\left[\epsilon^a N^b {\cal F}^i_{ab}\right]
    \nonumber\\
    \!\!&&\!\!
    - L_i\left[\epsilon^a N^b F^{0i}_{ab}\right]
    \,,\\
    \label{eq:HHa-Poisson}
    \{H[N],H_c[\epsilon^c]\}
    \!\!&=&\!\! - H[\mathcal{L}_{\vec \epsilon} N]
    + G_i\left[N\epsilon^a {\cal F}^i_{\bar{0}a}\right]
    \nonumber\\
    \!\!&&\!\!
    + L_i\left[N\epsilon^a F^{0i}_{\bar{0}a}\right]
    \,,\\
    \label{eq:HH-Poisson}
    \{H[N],H[\epsilon^{\bar{0}}]\}
    \!\!&=&\!\! - H_a\left[q^{ab} \left(\epsilon^{\bar{0}}\partial_bN-N\partial_b \epsilon^{\bar{0}}\right)\right]
    ,\qquad\\
    \label{eq:HaJ-Poisson}
    \{H_c[N^c],G_k[\theta^k]\}
    \!\!&=&\!\! 0
    \,,\\
    \label{eq:HaL-Poisson}
    \{H_c[N^c],L_k[\beta^k]\}
    \!\!&=&\!\! 0
    \,,\\
    \label{eq:HJ-Poisson}
    \{H[N],G_k[\theta^k]\}
    \!\!&=&\!\! 0
    \,,\\
    \label{eq:HL-Poisson}
    \{H[N],L_k[\beta^k]\}
    \!\!&=&\!\! 0
    \,,\\
    \label{eq:JJ-Poisson}
    \{G_i[\Gamma_t^i],G_j[\theta^j]\} \!\!&=&\!\! G_k \left[\tensor{\epsilon}{^k_i_j} \Gamma_t^i \theta^j\right]
    \,,\\
    \label{eq:JL-Poisson}
    \{G_i[\Gamma_t^i],L_j[\beta^j]\}
    \!\!&=&\!\! L_k \left[\tensor{\epsilon}{^k_i_j} \Gamma_t^i \beta^j\right]
    \,,\\
    \label{eq:LL-Poisson}
    \{L_i[K_t^i],L_j[\beta^j]\}
    \!\!&=&\!\! - G_k \left[\tensor{\epsilon}{^k_i_j} K_t^i \beta^j\right]
    \,,
\end{eqnarray}
which is identical to that of the vacuum system \cite{Tetrads}.

\subsection{Gauge transformations and covariance}

While the transformation of phase-space functions under the flow of the first-class constraints is directly provided by the action of Poisson brackets, this is not the case for the non-dynamical fields $N$, $N^a$, $K_t^i$, and $\Gamma_t^i$.
Instead, their transformations are defined by the preservation of the form of Hamilton's equations of motion along this flow---this is the definition of a gauge transformation---and hence are determined by the constraint algebra (\ref{eq:HaHa-Poisson})-(\ref{eq:LL-Poisson}) \cite{pons1997gauge,salisbury1983realization,Tetrads}:
\begin{eqnarray}\label{eq:Gauge transf - N}
    \delta_{\epsilon,\theta,\beta} N \!\!&=&\!\! \dot{\epsilon}^0 + \epsilon^b\partial_b N - N^b \partial_b \epsilon^{\bar{0}}
    \,,\\
    \delta_{\epsilon,\theta,\beta} N^a \!\!&=&\!\! \dot{\epsilon}^a + \epsilon^b\partial_b N^a - N^b \partial_b \epsilon^a
    \nonumber\\
    \!\!&&\!\!
    + q^{ab} \left(\epsilon^{\bar{0}}\partial_b N - N \partial_b \epsilon^{\bar{0}}\right)
    \,,\label{eq:Gauge transf - Na}\\
    \delta_{\epsilon,\theta,\beta} K_t^i \!\!&=&\!\! \dot{\beta}^i
    + \tensor{\epsilon}{^i_j_k} \left(\beta^j \Gamma_t^k+\theta^j K_t^k\right)
    + \epsilon^a N^b F^{0i}_{ab}
    \nonumber\\
    \!\!&&\!\!
    + \left(\epsilon^{\bar{0}}N^a-N\epsilon^a\right) F^{0i}_{\bar{0}a}
    \,,\label{eq:Gauge transf - Kt}\\
    \delta_{\epsilon,\theta,\beta} \Gamma_t^i \!\!&=&\!\! \dot{\theta}^i
    + \tensor{\epsilon}{^i_j_k} \left(\theta^j \Gamma_t^k
    - \beta^j K_t^k\right)
    + \epsilon^a N^b {\cal F}^i_{ab}
    \nonumber\\
    \!\!&&\!\!
    + \left(\epsilon^{\bar{0}}N^a-N\epsilon^a\right) {\cal F}^i_{\bar{0}a}
    \,.\label{eq:Gauge transf - Gammat}
\end{eqnarray}
That is, the canonical gauge transformations are given by the combined operation
\begin{eqnarray}
    {\cal O} \!\!&\to&\!\! {\cal O} + \{{\cal O},H[\epsilon^0]+H_a[\epsilon^a]+G_i[\theta^i]+L_i[\beta^i]\}
    \,,\nonumber\\
    N \!\!&\to&\!\! N + \delta_{\epsilon,\theta,\beta} N
    \quad,\quad
    N^a \to N^a + \delta_{\epsilon,\theta,\beta} N^a
    \,,\nonumber\\
    K_t^i \!\!&\to&\!\! K_t^i + \delta_{\epsilon,\theta,\beta} K_t^i
    \quad,\quad
    \Gamma_t^i \to \Gamma_t^i + \delta_{\epsilon,\theta,\beta} \Gamma_t^i
    \,,
\end{eqnarray}
for all phase-space functions ${\cal O}$.

The transformations of the non-dynamical fields are essential to understand the transformations of the physical fields which generally depend on them---for  instance, the components of the spacetime metric (\ref{eq:ADM line element}) depend on $N$ and $N^a$, while the connection, and hence also the strength and torsion tensor fields, depend on $K_t^i$ and $\Gamma_t^i$.

The canonical system is covariant only if the canonical gauge transformations of the physical fields correspond to spacetime diffeomorphisms and infinitesimal SL(2,$\mathbb{C}$) transformations on shell.
For instance, the canonical gauge transformation of the spacetime metric must satisfy the covariance condition
\begin{equation}
    \delta_{\epsilon,\theta,\beta} g_{\mu \nu} \big|_{\rm OS} =
    \mathcal{L}_{\xi} g_{\mu \nu} \big|_{\rm OS}
    \,,
    \label{eq:Spacetime covariance condition}
\end{equation}
where "OS" denotes an evaluation on shell, involving the vanishing of the constraints and use of Hamilton's equations of motion in the right-hand side for the time-derivatives, $\dot{q}_{ab}=\{q_{ab},{\cal H}\}$.
The gauge parameters $(\epsilon^{\bar{0}}, \epsilon^a)$ in (\ref{eq:Spacetime covariance condition}) are related to the diffeomorphism-generator vector field $\xi^\mu$ by the following change of basis,
\begin{eqnarray}
    \xi^\mu \!\!&=&\!\! \epsilon^{\bar{0}} n^\mu + \epsilon^a s^\mu_a
    = \xi^t t^\mu + \xi^a s^\mu_a
    \,,\nonumber
    \\
    \xi^t \!\!&=&\!\! \frac{\epsilon^{\bar{0}}}{N}
    \quad,\quad
    \xi^a = \epsilon^a - \frac{\epsilon^{\bar{0}}}{N} N^a
    \,.
\label{eq:Diffeomorphism generator projection}
\end{eqnarray}

On the other hand, the canonical decomposition of spacetime diffeomorphisms and proper Lorentz transformations---the latter denoted by $\delta^{\rm SL(2,\mathbb{C})}$ in what follows---of the non-dynamical components of the connection are respectively given by
\begin{eqnarray}
    \mathcal{L}_\xi K_t^i
    \!\!&=&\!\! \delta^{\rm SL(2,\mathbb{C})}_{\xi^\mu \Gamma_\mu,\xi^\mu K_\mu} K_t^i
    + \epsilon^a N^b F_{ab}^{0i}
    \nonumber\\
    \!\!&&\!\!
    + \left(\epsilon^{\bar{0}}N^a-N\epsilon^a\right) F_{0 a}^{0i}
    \,,\\
    \mathcal{L}_\xi \Gamma_t^i \!\!&=&\!\! \delta^{\rm SL(2,\mathbb{C})}_{\xi^\mu \Gamma_\mu,\xi^\mu K_\mu} \Gamma_t^i
    + \epsilon^a N^b {\cal F}_{ab}^i
    \nonumber\\
    \!\!&&\!\!
    + \left(\epsilon^{\bar{0}}N^a-N\epsilon^a\right) {\cal F}_{0 a}^i
    \,,
\end{eqnarray}
and
\begin{eqnarray}
    \delta^{\rm SL(2,\mathbb{C})}_{\theta,\beta} K_t^i \!\!&=&\!\! \dot{\beta}^i
    + \tensor{\epsilon}{^i_j_k} \left( \beta^j \Gamma_t^k
    + \theta^j K_t^k\right)
    \,,\\
    \delta^{\rm SL(2,\mathbb{C})}_{\theta,\beta} \Gamma_t^i \!\!&=&\!\!
    \dot{\theta}^i
    + \tensor{\epsilon}{^i_j_k} \left(\theta^j \Gamma_t^k - \beta^j K_t^k\right)\,.
\end{eqnarray}
Using this, the canonical gauge transformations (\ref{eq:Gauge transf - N})-(\ref{eq:Gauge transf - Gammat}) can be written as linear combinations of diffeomorphisms and Lorentz transformations:
\begin{equation}\label{eq:Canonical gauge transformation}
    \delta_{\epsilon,\theta,\beta} \cdot \big|_{\rm OS} = \mathcal{L}_{\xi} \cdot + \delta^{\rm SL(2,\mathbb{C})}_{\theta-\xi^\mu \Gamma_\mu,\beta-\xi^\mu K_\mu} \cdot \big|_{\rm OS}\,.
\end{equation}
All physical fields are therefore expected to follow this transformation under the gauge flow.

In particular, the covariance condition of the spacetime metric (\ref{eq:Spacetime covariance condition}) agrees with the transformation (\ref{eq:Canonical gauge transformation}) if the metric is Lorentz-invariant under the gauge flow.
This is indeed the case because the lapse and shift do not suffer transformations under the flow of the Lorentz--Gauss constraints, as implied by (\ref{eq:Gauge transf - N}) and (\ref{eq:Gauge transf - Na}), and neither does the spatial metric,
\begin{eqnarray}\label{eq:qG}
    \{q^{ab} , G_i[\theta^i]\} \!\!&=&\!\! 0
    \,,\\
    \{q^{ab} , L_i[\beta^i]\} \!\!&=&\!\! 0
    \,.
\end{eqnarray}
This, in turn, implies that the vector constraint generates spatial diffeomorphisms of the spatial metric on shell:
\begin{eqnarray}
    \{q^{ab}(x),H_c[\epsilon^c]\} \!\!&=&\!\! \mathcal{L}_{\vec \epsilon}\, q^{ab}
    - L_i\left[\{q^{ab}(x),K_b^iN^b\}\right]
    \nonumber\\
    \!\!&&\!\!
    - G_i\left[\{q^{ab}(x),\Gamma_b^iN^b\}\right]
    \,.
\end{eqnarray}
Finally, the canonical gauge transformation of the spatial metric does not contain derivatives of the normal gauge function,
\begin{equation}\label{eq:Spacetime covariance condition - reduced}
    \frac{\partial \{q^{ab},H[\epsilon^{\bar{0}}]\}}{\partial (\partial_{c_1} \epsilon^{\bar{0}})} \bigg|_{\rm OS} = \frac{\partial \{q^{ab},H[\epsilon^{\bar{0}}]\}}{\partial (\partial_{c_1} \partial_{c_2} \epsilon^{\bar{0}})} \bigg|_{\rm OS} = \dotsi = 0\,.
\end{equation}
This is a necessary condition---not implied by the constraint algebra---for (\ref{eq:Spacetime covariance condition}) to hold \cite{EMGCov}.
Using equations (\ref{eq:Gauge transf - N}), (\ref{eq:Gauge transf - Na}), and (\ref{eq:qG})-(\ref{eq:Spacetime covariance condition - reduced}), we confirm that the covariance condition (\ref{eq:Spacetime covariance condition}) is indeed satisfied.

Using (\ref{eq:Gauge transf - N}), (\ref{eq:Gauge transf - Na}), and the fact that the Lorentz, Gauss, and vector constraints generate a linear combination of Lorentz transformations and spatial diffeomorphisms, the covariance condition of the tetrad
\begin{equation}
    \delta_{\epsilon,\theta,\beta} e_\mu^I \big|_{\rm OS} =
    \mathcal{L}_{\xi} e_\mu^I + \delta^{\rm SL(2,\mathbb{C})}_{\theta-\xi^\mu \Gamma_\mu,\beta-\xi^\mu K_\mu} e_\mu^I \big|_{\rm OS}\,,
    \label{eq:Tetrad covariance condition}
\end{equation}
can be reduced to a set of conditions for the Hamiltonian constraint, given by
\begin{eqnarray}
    \frac{\partial \{v_i,H[\epsilon^{\bar{0}}]\}}{\partial(\partial_c\epsilon^{\bar{0}})} \bigg|_{\rm O.S.} 
    \!\!&=&\!\! \frac{\sqrt{(\det {\cal P})^{-1}}}{\gamma\sqrt{8\pi G}} {\cal P}^c_i \bigg|_{\rm OS} 
    \,,\label{eq:Tetrad covariance condition - reduced - 1}\\
    \frac{\partial^2 \{v_i,H[\epsilon^{\bar{0}}]\}}{\partial(\partial_{c_1}\partial_{c_2}\epsilon^{\bar{0}})} \bigg|_{\rm OS} 
    \!\!&=&\!\! 0
    \,,\label{eq:Tetrad covariance condition - reduced - 2}
\end{eqnarray}
and
\begin{eqnarray}
    \!\!&&\!\!
    \left(\delta^a_c\delta^k_i
    - \frac{1}{2} {\cal P}^a_i ({\cal P}^{-1})^k_c \right) \frac{\partial\{{\cal P}^c_k,H[\epsilon^{\bar{0}}]\}}{\partial(\partial_d \epsilon^{\bar{0}})} \bigg|_{\rm OS}
    \nonumber\\
    \!\!&&\!\!\qquad=
    - \frac{\gamma \sqrt{(\det {\cal P})^{-1}}}{\sqrt{8\pi G}} v^j {\cal P}^a_j {\cal P}^d_i \bigg|_{\rm OS}
    \,,\label{eq:Tetrad covariance condition - reduced - 3}\\
    \!\!&&\!\!
    \left(\delta^a_c\delta^k_i
    - \frac{1}{2} {\cal P}^a_i ({\cal P}^{-1})^k_c \right) \frac{\partial^2\{{\cal P}^c_k,H[\epsilon^{\bar{0}}]\}}{\partial(\partial_{d_1} \partial_{d_2} \epsilon^{\bar{0}})} \bigg|_{\rm OS} = 0
    \,.\qquad
    \label{eq:Tetrad covariance condition - reduced - 4}
\end{eqnarray}
Similarly, the covariance condition of the connection,
\begin{equation}
    \delta_{\epsilon,\theta,\beta} \tensor{\omega}{_\mu^I^J} \big|_{\rm OS} =
    \mathcal{L}_{\xi} \tensor{\omega}{_\mu^I^J} + \delta^{\rm SL(2,\mathbb{C})}_{\theta-\xi^\mu \Gamma_\mu,\beta-\xi^\mu K_\mu} \tensor{\omega}{_\mu^I^J} \big|_{\rm OS}
    \,,
    \label{eq:Connection covariance condition}
\end{equation}
reduces to
\begin{equation}\label{eq:Connection covariance condition - reduced}
    \frac{\partial \{\tensor{\omega}{_a^I^J},H[\epsilon^{\bar{0}}]\}}{\partial (\partial_{c_1} \epsilon^{\bar{0}})} \bigg|_{\rm OS} = \frac{\partial \{\tensor{\omega}{_a^I^J},H[\epsilon^{\bar{0}}]\}}{\partial (\partial_{c_1} \partial_{c_2} \epsilon^{\bar{0}})} \bigg|_{\rm OS} = \dotsi = 0\,,
\end{equation}
that of the fermion field,
\begin{equation}
    \delta_{\epsilon,\theta,\beta} \Psi \big|_{\rm OS} =
    \mathcal{L}_{\xi} \Psi + \delta^{\rm SL(2,\mathbb{C})}_{\theta-\xi^\mu \Gamma_\mu,\beta-\xi^\mu K_\mu} \Psi \big|_{\rm OS}\,,
    \label{eq:Spinor covariance condition}
\end{equation}
to
\begin{equation}\label{eq:Spinor covariance condition - reduced}
    \frac{\partial \{\Psi,H[\epsilon^{\bar{0}}]\}}{\partial (\partial_{c_1} \epsilon^{\bar{0}})} \bigg|_{\rm OS} = \frac{\partial \{\Psi,H[\epsilon^{\bar{0}}]\}}{\partial (\partial_{c_1} \partial_{c_2} \epsilon^{\bar{0}})} \bigg|_{\rm OS} = \dotsi = 0\,,
\end{equation}
and that of the fermion momentum,
\begin{equation}
    \delta_{\epsilon,\theta,\beta} P_\Psi \big|_{\rm OS} =
    \mathcal{L}_{\xi} P_\Psi + \delta^{\rm SL(2,\mathbb{C})}_{\theta-\xi^\mu \Gamma_\mu,\beta-\xi^\mu K_\mu} P_\Psi \big|_{\rm OS}\,,
    \label{eq:Spinor momentum covariance condition}
\end{equation}
to
\begin{equation}\label{eq:Spinor momentum covariance condition - reduced}
    \frac{\partial \{P_\Psi \overline{\Xi^{-1}},H[\epsilon^{\bar{0}}]\}}{\partial (\partial_{c_1} \epsilon^{\bar{0}})} \bigg|_{\rm OS} = \frac{\partial \{P_\Psi \overline{\Xi^{-1}},H[\epsilon^{\bar{0}}]\}}{\partial (\partial_{c_1} \partial_{c_2} \epsilon^{\bar{0}})} \bigg|_{\rm OS} = \dotsi = 0\,,
\end{equation}
or, using (\ref{eq:Spacetime covariance condition - reduced}) and (\ref{eq:Tetrad covariance condition - reduced - 1}), to
\begin{equation}\label{eq:Spinor momentum covariance condition - reduced - 1}
    \frac{\partial \{ P_\Psi ,H[\epsilon^{\bar{0}}]\}}{\partial (\partial_{c} \epsilon^{\bar{0}})}
    \bigg|_{\rm OS}
    = \frac{8\pi G}{\sqrt{\det q}} {\cal P}^c_j P_\Psi \left[ v^j - \left(\gamma^0
    - v_i \gamma^i\right) \gamma^j\right]
\end{equation}
and
\begin{equation}\label{eq:Spinor momentum covariance condition - reduced - 2}
    \frac{\partial \{P_\Psi,H[\epsilon^{\bar{0}}]\}}{\partial (\partial_{c_1} \partial_{c_2} \epsilon^{\bar{0}})} \bigg|_{\rm OS} = 0\,,
\end{equation}
where we used the fact that $P_\Psi\overline{\Xi^{-1}}$ constitutes a co-spinor field to arrive at (\ref{eq:Spinor momentum covariance condition - reduced}).
The covariance conditions of $\overline{\Psi}$ and $\overline{P_\Psi}$ are given by the complex conjugate of those of $\Psi$ and $P_\Psi$ because all the constraints are Hermitian upon use of the reality conditions and hence do not constitute independent covariance conditions.

The connection and fermion covariance conditions (\ref{eq:Connection covariance condition - reduced}) and (\ref{eq:Spinor covariance condition - reduced}) can be readily verified to hold because the Hamiltonian constraint has no derivatives of their momenta; similarly, the covariance conditions for the fermion momentum (\ref{eq:Spinor momentum covariance condition - reduced - 1}) and (\ref{eq:Spinor momentum covariance condition - reduced - 2}) can be easily shown to hold from the simple dependence of the $H$ on derivatives of $\Psi$.
On the other hand, proving that the covariance conditions of the tetrad (\ref{eq:Tetrad covariance condition - reduced - 1})-(\ref{eq:Tetrad covariance condition - reduced - 4}) hold requires several detailed computations, but they indeed hold in the vacuum \cite{Tetrads} and the presence of fermions does not imply contributions to these equations because the respective contributions to the constraints do not contain derivatives of the configuration variables conjugate to $v_i$ or ${\cal P}^a_i$.

All physical fields correspond to geometric objects, matter currents, or stress-energy tensor fields that are constructed with the tetrad, the connection, and $\Psi$; therefore, the canonical gauge transformations of all physical fields correspond to linear combinations of spacetime diffeomorphisms and ${\rm SL}(2,\mathbb{C})$ transformations.
We conclude that our canonical system is indeed covariant in the extended phase space.
This property holds on the second-class constraint surface, which we show in the next section by use of the corresponding Dirac brackets.

\subsection{Dirac observables}

The functionals
\begin{eqnarray}\label{eq:Lorentz observable}
    \mathfrak{L}_i[\alpha^i] \!\!&=&\!\!
    \int{\rm d}^3x \alpha^i \bigg[- \left( \tilde{\cal P}^a_k \Gamma^l_a
    - \tilde{\cal K}^a_k K^l_a \right)\tensor{\epsilon}{^k_l_i}
    \\
    \!\!&&\!\!\qquad\qquad
    - i \left(P_\Psi S_{0i} {\Psi} - \overline{\Psi}\, S_{0i} \overline{P_\Psi} \right)\bigg]
    \,,\nonumber
\end{eqnarray}
and
\begin{eqnarray}\label{eq:Gauss observable}
    \mathfrak{G}_i [\sigma^i] \!\!&=&\!\!
    \int{\rm d}^3x \sigma^i \bigg[- \left( \tilde{\cal P}^a_k K^l_a
    + \tilde{\cal K}^a_k \Gamma^l_a \right) \tensor{\epsilon}{_i_k^l}
    \\
    \!\!&&\!\!\qquad\qquad
    - i \left[P_\Psi S_i {\Psi} - \overline{\Psi}\,S_i \overline{P_\Psi} \right]
    \bigg]
    \,,\nonumber
\end{eqnarray}
with constants $\alpha^i$ and $\sigma^i$, are (nonlocal) Dirac observables because they are identical to the Lorentz--Gauss constraints up to boundary terms when smeared by constants: Using $\{\mathfrak{L}_i,{\cal H}\}=\{L_i+\partial_a\tilde{\cal P}^a_i,{\cal H}\}$, and $\{\mathfrak{G}_i,{\cal H}\}=\{G_i+\partial_a\tilde{\cal K}^a_i,{\cal H}\}$, the brackets of the local versions of the observables with the constraints result in
\begin{equation}
    \{\mathfrak{L}_i,L_j[K_t^j]\} |_{\rm OS} =
    %\tensor{\epsilon}{_i_j^k} K_t^j G_k +
    \partial_a \left(\tensor{\epsilon}{_i_k^l} K_t^k\tilde{\cal K}^a_l\right)
    \,,
\end{equation}
\begin{equation}
    \{\mathfrak{L}_i,G_j[\Gamma_t^j]\} |_{\rm OS} =
    %- \tensor{\epsilon}{_i_j^k} \Gamma_t^j L_k
    - \partial_a \left( \tensor{\epsilon}{_i_k^l} \Gamma_t^k \tilde{\cal P}^a_l\right)
    \,,
\end{equation}
\begin{eqnarray}\label{eq:LHa obs}
    \{\mathfrak{L}_i,H_a[N^a]\} |_{\rm OS} \!\!&=&\!\! \partial_a\left( \mathfrak{L}_i N^a\right)
    \\
    \!\!&&\!\!
    + \partial_a\left(\tensor{\epsilon}{_i_j^k} \left(\Gamma^j_b \tilde{\cal P}^a_k-K^j_b \tilde{\cal K}^a_k\right) N^b\right)
    \nonumber
    \,,
\end{eqnarray}
and
\begin{equation}
    \{\mathfrak{G}_i,L_j[K_t^j]\} |_{\rm OS} = %- \tensor{\epsilon}{_i_j^k} K_t^j L_k
    - \partial_a \left( \tensor{\epsilon}{_i_k^l} K_t^k \tilde{\cal P}^a_l\right)
    \,,
\end{equation}
\begin{equation}
    \{\mathfrak{G}_i,G_j[\Gamma_t^j]\} |_{\rm OS} = %- \tensor{\epsilon}{_i_j^k} \Gamma_t^j G_k
    - \partial_a \left(\tensor{\epsilon}{_i_k^l} \Gamma_t^k \tilde{\cal K}^a_l\right)
    \,.
\end{equation}
\begin{eqnarray}
    \{\mathfrak{G}_i,H_a[N^a]\} |_{\rm OS} \!\!&&\!\! \partial_a \left( \mathfrak{G}_i N^a\right)
    \\
    \!\!&&\!\!
    + \partial_a\left(\tensor{\epsilon}{_i_k^l} \left(K^k_b \tilde{\cal P}^a_l + \Gamma^k_b \tilde{\cal K}^a_l \right) N^b\right)
    \,,\nonumber
\end{eqnarray}
as well as
\begin{widetext}
\begin{eqnarray}
    \{\mathfrak{L}_i,H[N]\} |_{\rm OS}
    \!\!&=&\!\! \partial_a \Bigg[N \frac{\gamma \sqrt{(\det {\cal P})^{-1}}}{\sqrt{8\pi G}} \bigg(
    \mathfrak{L}_i v^q {\cal P}^a_q
    + \tensor{\epsilon}{_i_j^k} \left(K^j_b \tilde{\cal K}^a_k 
    - \Gamma^j_b \tilde{\cal P}^a_k\right) v^q {\cal P}^b_q
    + \frac{2}{\gamma^2} {\cal P}^{[a}_i {\cal P}^{b]}_q \left(K_b^q - \zeta \Gamma_b^q\right)
    \nonumber\\
    \!\!&&\!\!\qquad\qquad\qquad\qquad\quad
    + \left({\cal P}^a_j i P_\Psi \left(\gamma^0-v_m\gamma^m\right) \gamma^j S_{0i} \Psi
    + c.c. \right)
    \bigg)
    \Bigg]
    \,,
\end{eqnarray}
\begin{eqnarray}
    \{\mathfrak{G}_i,H[N]\} |_{\rm OS} \!\!&=&\!\! \partial_a \Bigg[N \frac{\gamma \sqrt{(\det {\cal P})^{-1}}}{\sqrt{8\pi G}} \bigg(
    \mathfrak{G}_i v^q {\cal P}^a_q
    + \tensor{\epsilon}{_i_k^l} \left(K^k_b \tilde{\cal P}^a_l + \Gamma^k_b \tilde{\cal K}^a_l \right) v^q {\cal P}^b_q
    - \frac{{\cal P}^{[a}_i {\cal P}^{b]}_q}{\gamma^2} \left(\Gamma_b^q
    + \zeta K_b^q\right)
    \nonumber\\
    \!\!&&\!\!\qquad\qquad\qquad\qquad\quad
    + \left({\cal P}^a_j P_\Psi \left(\gamma^0-v_m\gamma^m\right) \gamma^j i S_i \Psi
    + c.c.\right)
    \bigg)
    \Bigg]
    \,.\label{eq:JH obs}
\end{eqnarray}
\end{widetext}

The boundary terms imply conserved densitized currents: Defining
\begin{equation}
    \mathcal{L}_i^t \equiv \mathfrak{L}_i
    \quad,\quad
    \{\mathfrak{L}_i,{\cal H}\} |_{\rm OS} =: - \partial_a \mathfrak{L}_i^a
    \,,
\end{equation}
and
\begin{equation}
    \mathcal{G}_i^t \equiv \mathfrak{G}_i
    \quad,\quad
    \{\mathfrak{G}_i,{\cal H}\} |_{\rm OS} =: - \partial_a \mathfrak{G}_i^a
    \,,
\end{equation}
the densitized four-currents ${\cal L}_i^\mu=({\cal L}_i^t,{\cal L}_i^a)$ and ${\cal G}_i^\mu=({\cal G}_i^t,{\cal G}_i^a)$ are conserved on shell by definition: $\dot{\cal L}_i^t=\{{\cal L}_i,{\cal H}\}|_{\rm OS}=-\partial_a{\cal L}_i^a$ and $\dot{\cal G}_i^t=\{{\cal G}_i,{\cal H}\}|_{\rm OS}=-\partial_a{\cal G}_i^a$, and hence $\partial_\mu{\cal L}_i^\mu=0$ and $\partial_\mu{\cal G}_i^\mu=0$.

Moreover, these observables form a local Lorentz algebra,
\begin{eqnarray}\label{eq:JJ - obs}
    \{\mathfrak{G}_i(x),\mathfrak{G}_j(y)\}
    \!\!&=&\!\! \tensor{\epsilon}{_i_j^k} \mathfrak{G}_k \delta^3(x-y)\,,
    \\
    \label{eq:JL - obs}
    \{\mathfrak{G}_i(x),\mathfrak{L}_j(y)\}
    \!\!&=&\!\! \tensor{\epsilon}{_i_j^k} \mathfrak{L}_k \delta^3(x-y)\,,
    \\
    \label{eq:LL - obs}
    \{\mathfrak{L}_i(x),\mathfrak{L}_j(y)\}
    \!\!&=&\!\! - \tensor{\epsilon}{_i_j^k} \mathfrak{G}_k \delta^3(x-y)\,.
\end{eqnarray}

The appearance of the fermionic variables in the observables (\ref{eq:Lorentz observable}) and (\ref{eq:Gauss observable}) clarifies their physical meaning: They are spin charges associated to the conserved spin currents $\mathcal{L}_i^\mu$ and $\mathcal{G}_i^\mu$.

Similarly, we expect that the densitized electric current $J^\mu=|e| e^\mu_I \overline{\Psi} \gamma^I \Psi$ be conserved too.
Indeed, the phase-space functional
\begin{equation}\label{eq:Electric charge observable}
    {\cal Q} [\alpha] = - i \int{\rm d}^3x \alpha \left(P_\Psi \Psi - \overline{\Psi}\, \overline{P_\Psi}\right)
    \,,
\end{equation}
whose local version is equivalent to $J^t$ if the relation (\ref{eq:Momenta Dirac fermion}) is used, commutes with the first-class constraints up to boundary terms:
\begin{equation}
    \{{\cal Q},L_j[K_t^j]\} =
    0
    \,,
\end{equation}
\begin{equation}
    \{{\cal Q},G_j[\Gamma_t^j]\} = 0
    \,,
\end{equation}
\begin{equation}
    \{{\cal Q},H_a[N^a]\} = \partial_a\left( \mathcal{Q} N^a\right)
    \,,
\end{equation}
and
\begin{widetext}
\begin{eqnarray}
    \{{\cal Q},H[N]\} \!\!&=&\!\! \partial_a \left[N \frac{\gamma \sqrt{(\det {\cal P})^{-1}}}{\sqrt{8\pi G}} {\cal P}^a_j \left(\mathcal{Q} v^j
    + i P_\Psi \left(\gamma^0-v_m\gamma^m\right) \gamma^j \Psi
    - i \overline{\Psi} \gamma^j \left(\gamma^0-v_m\gamma^m\right) \overline{P_\Psi} \right)
    \right]
    \,.
\end{eqnarray}
\end{widetext}
The boundary terms can be identified with the spatial components of the electric current $J^a$ such that
\begin{equation}
    \dot{\cal Q}=-\partial_a J^a
\end{equation}
or $\partial_\mu J^\mu=0$ on shell.

The bracket of this observable with itself forms a local Abelian algebra,
\begin{equation}
    \{{\cal Q}(x),{\cal Q}(y)\}=0\,.
\end{equation}

\subsection{Fermion dynamics in the extended phase space}
\label{sec:Fermion dynamics in the extended phase space}

Hamilton's equations of motion for the fermionic variables result in
\begin{equation}\label{eq:Psi dot}
    \dot{\Psi} = \{\Psi,{\cal H}\}
    = \Theta^a \partial_a \Psi
    + i \Omega \Psi
    \,,
\end{equation}
and
\begin{equation}
    \label{eq:P_Psi dot}
    \dot{P}_\Psi = \{P_\Psi,{\cal H}\} = \partial_a \left(P_\Psi\Theta^a\right)
    - i P_\Psi \Omega
    \,.
\end{equation}
A direct inspection shows that (\ref{eq:Psi dot}) is equivalent to (\ref{eq:Extended fermion EoM Psi}).

Furthermore, using (\ref{eq:Spinor momentum Phi}), the equation of motion of the momentum (\ref{eq:P_Psi dot}) can be written as
\begin{eqnarray}\label{eq:Phi dot}
    \dot{\Phi} \!\!&=&\!\! \Theta^a \partial_a \Phi
    + i \Omega \Phi
    \\
    \!\!&&\!\!
    - \Xi^{-1} \left[\{\Xi,{\cal H}\}
    - \partial_a \left(\overline{\Theta^a} \Xi\right)
    - i \left(\overline{\Omega}\Xi-\Xi \Omega\right)\right] \Phi
    \,,\nonumber
\end{eqnarray}
upon complex conjugation, where the last line is proportional to the covariant divergence of the tetrad (\ref{eq:Covariant divergence tetrad - phase space}) and hence (\ref{eq:Phi dot})
is identical to the equation of motion (\ref{eq:Extended fermion EoM Phi}).
This confirms that the canonical system, off the Dirac-fermion constraint surface, indeed describes the dynamics of the extended fermion action (\ref{eq:Extended Dirac action}).

For completeness, we compute
\begin{widetext}
\begin{eqnarray}
    \!\!&&\!\!
    \{\Xi,{\cal H}\}
    - \partial_a \left(\overline{\Theta^a} \Xi\right)
    - i \left(\overline{\Omega}\Xi-\Xi \Omega\right) |_{\rm OS}
    \\
    \!\!&&\!\!\qquad = \frac{3i}{8} \frac{8\pi G \gamma N}{1+\zeta^2} \left(1-i\alpha\gamma^5\right) \Bigg[ P_\Psi \left(\gamma^0-v_m\gamma^m\right) \gamma^0 \left(1 + \zeta i \gamma^5\right) \Psi \gamma^0
    - P_\Psi \left(\gamma^0-v_s\gamma^s\right) \gamma_i \left(1 + \zeta i \gamma^5\right) \Psi \gamma^i
    \nonumber\\
    \!\!&&\!\!\qquad\qquad\qquad\qquad\qquad\qquad\quad
    - \frac{2}{3} \delta_{i[p} \mathfrak{K}^p_{l]} P_\Psi \left(\gamma^0-v_s\gamma^s\right) \gamma^l \left(i \gamma^5 - \zeta\right) \Psi \gamma^i + c.c.\Bigg]
    \nonumber\\
    \!\!&&\!\!\qquad
    = \frac{3i}{16} \frac{\sqrt{\det q}}{\gamma} \frac{8\pi G N}{1+\zeta^2} \left(1-i\alpha\gamma^5\right) \Bigg[- \overline{\Phi} \gamma^0 \left[\left(\zeta-\alpha\right) \gamma^5 - i \left(1 + \alpha \zeta\right)\right] \Psi \gamma^0
    + \overline{\Phi} \gamma_i \left[\left(\zeta-\alpha\right) \gamma^5-i \left(1+\alpha \zeta\right)\right] \Psi \gamma^i
    \nonumber\\
    \!\!&&\!\!\qquad\qquad\qquad\qquad\qquad\qquad\qquad\qquad
    + \frac{2}{3} \delta_{i[p} \mathfrak{K}^p_{l]} \overline{\Phi} \gamma^l \left( \left(1+\zeta\alpha\right) \gamma^5-i\left(\alpha-\zeta\right) \right) \Psi \gamma^i
    + c.c \Bigg]
    \,,\nonumber
\end{eqnarray}
\end{widetext}
where ${\rm OS}$ here refers to an evaluation on the first-class constraint surface and we used (\ref{eq:Spinor momentum Phi}) to get the second equality---see App.~\ref{app:Hamiltonian constraint} for intermediate steps of this calculation.

It follows that the Dirac-fermion constraint (\ref{eq:Dirac fermion constraint - 2})---and hence also (\ref{eq:Dirac fermion constraint - 1})---is second-class for non-trivial $\zeta$, $\alpha$, or $\mathfrak{K}_{ij}$:
\begin{widetext}
\begin{eqnarray}\label{eq:C_Psi,H}
    \{C_\Psi,{\cal H}\} |_{\rm OS} \!\!&=&\!\! \left[\{\Xi,{\cal H}\}
    - \partial_a \left(\overline{\Theta^a} \Xi\right)
    - i \left(\overline{\Omega} \Xi-\Xi \Omega\right)\right] \Psi
    \\
    \!\!&=&\!\! \frac{3 i}{8} \frac{\sqrt{\det q}}{\gamma} \frac{8\pi G N}{1+\zeta^2} \left(1-i\alpha\gamma^5\right) \left[- \overline{\Psi} \gamma^0 \left(\zeta-\alpha\right) \gamma^5 \Psi \gamma^0
    + \overline{\Psi} \gamma^l \left(\delta_{li} \left(\zeta-\alpha\right)
    + \frac{2}{3} \delta_{i[p} \mathfrak{K}^p_{l]} \left(1+\zeta\alpha\right)\right) \gamma^5 \Psi \gamma^i
    \right] \Psi
    \,,\nonumber
\end{eqnarray}
\end{widetext}
where we set $C_\Psi=0=C_{\overline{\Psi}}$ after the evaluation of the brackets.

\section{Second-class constraints}
\label{sec:Second-class constraints}

\subsection{Gravitational second-class constraints}

The fermionic contributions to the Lorentz, Gauss, and vector constraints do not involve the components of the connection in a way that they can contribute to the gravitational secondary second-class constraint (\ref{eq:Second-class constraint B=0}) on shell, while their gravitational contributions similarly vanish on shell upon evaluation of the primary second-class constraint (\ref{eq:Second-class constraint K=0}) \cite{Tetrads}.
Therefore, the only nontrivial contribution comes from the second term of the Hamiltonian constraint,
\begin{equation}
    \frac{\delta H^{(2)}[N]}{\delta {\cal B}^{ij}}
    = N \frac{\sqrt{(\det {\cal P})^{-1}}}{\gamma\sqrt{8\pi G}} {\cal T}_{ij}
    \,,\nonumber
\end{equation}
where ${\cal T}_{ij}$ is given by the torsion term (\ref{eq:Torsion-spatial-sym}).
This expression can be written as
\begin{equation}\label{eq:Second-class constraint}
    {\cal T}_{ij} = \frac{1}{1+\zeta^2} \tensor{V}{_i_j^k^l} \left({\cal B}_{kl}-\bar{\cal B}_{kl}\right)\,,
\end{equation}
such that ${\cal B}^{kl}=\bar{\cal B}^{kl}$ solves the second-class constraint ${\cal C}_{ij}=0$---implying that no tertiary second-class constraints arise---and where
\begin{equation}
    \tensor{V}{^k^l_i_j} = \left( \delta_{pq} - v_p v_q \right) \tensor{\epsilon}{^k^p_{(i}} \tensor{\epsilon}{^l^q_{j)}}
    =: V_{pq} \tensor{\epsilon}{^k^p_{(i}} \tensor{\epsilon}{^l^q_{j)}}\,,
\end{equation}
using $V_{mn} = \delta_{mn} - v_nv_m$, and
\begin{equation}
    \bar{\cal B}_{ij} = \tensor{\left(V^{-1}\right)}{_i_j^k^l} \mathfrak{b}_{kl}
    \,,
\end{equation}
with
\begin{eqnarray}
    \mathfrak{b}_{kl} \!\!&=&\!\! \left(\delta_{p(k} {\cal K}^d_{l)} 
    + \zeta \left(\delta^q_p\delta_{kl}-\delta_{p(k}\delta^q_{l)}\right) {\cal P}^d_q \right) {\cal D}_d^p
    \\
    \!\!&&\!\!
    - \zeta v_{(k} {\cal E}_{l)}
    - (1+\zeta^2) \delta_{m(k} \tensor{\epsilon}{_{l)}^p^q} \left({\cal P}^{-1}\right)_c^m {\cal P}_p^d \partial_d {\cal P}^c_q
    \nonumber\\
    \!\!&&\!\!
    - \frac{1}{2} \left[P_\Psi \left(\zeta - i \gamma^5\right) \upsilon_{kl} \Psi
    + \overline{\Psi}\,\overline{\upsilon_{kl}} \left(\zeta - i \gamma^5\right) \overline{P_\Psi}
    \right]
    \,,\nonumber
\end{eqnarray}
and
\begin{equation}
    \tensor{\left(V^{-1}\right)}{^r^s_k_l}
    = \frac{1}{2\gamma^2} \left(V^{rs} V_{kl} - 2 V^r_{(k}V^s_{l)}\right)
\end{equation}
is the inverse of $\tensor{V}{^k^l_i_j}$:
\begin{equation}
    \tensor{\left(V^{-1}\right)}{^r^s_k_l} \tensor{V}{^k^l_i_j} = \delta^{(r}_k\delta^{s)}_l\,.
\end{equation}

We can substitute the solution to the second-class constraints, ${\cal B}_{ij}=\bar{\cal B}_{ij}$ and $\mathfrak{K}_{ij}=0$, in the first-class constraints and the symplectic structure, thereby reducing the phase space, now spanned by the canonical pairs $({\cal P},{\cal D})$, $(v,{\cal E})$, $(\Psi,P_\Psi)$, and $(\overline{\Psi},\overline{P_\Psi})$.
However, the constraints turn out to have a complicated dependence on the reduced phase space, which complicates explicit computations of their brackets.
This complication is bypassed by preserving the extended phase space and use the corresponding Dirac brackets.

\subsection{Dirac brackets}
\label{sec:Dirac brackets}

The Poisson brackets between the second-class constraints in the extended phase space are given by
\begin{equation}
    \{\mathfrak{K}_{ij}(x),\mathfrak{K}^{kl}(y)\} = 0
    \,,
\end{equation}
\begin{equation}
    \{\mathfrak{K}_{ij}(x),{\cal T}^{kl}(y)\} = - \frac{1}{1+\zeta^2} \tensor{V}{_i_j^k^l} \delta^3(x-y)
    \,,
\end{equation}
\begin{equation}
    \{\mathfrak{K}_{ij}(x),C_{\overline{\Psi}}(y)\} = 0
    \quad,\quad
    \{\mathfrak{K}_{ij}(x),{C_\Psi}(y)\} = 0\,,
\end{equation}
\begin{equation}
    \{C_{\overline{\Psi}}(x),C_{\overline{\Psi}}(y)\} = 0
    \quad,\quad
    \{{C_\Psi}(x),{C_\Psi}(y)\} = 0
    \,,
\end{equation}
and
\begin{equation}
    \{C_{\overline{\Psi}}(x),{C_\Psi}(y)\} = \Upsilon \delta^3 (x-y)\,,
\end{equation}
where
\begin{equation}
    \Upsilon=i (8\pi G)^{3/2} \sqrt{|\det {\cal P}|} \left(\gamma^0 - v_m \gamma^m\right)\,,
\end{equation}
with inverse
\begin{equation}
    \Upsilon^{-1}= \frac{- i\gamma^2}{(8\pi G)^{3/2} \sqrt{|\det {\cal P}|}} \left(\gamma^0 - v_m \gamma^m\right)\,,
\end{equation}
\begin{equation}
    \{{\cal T}_{ij}(x),C_{\overline{\Psi}}(y)\}|_{\rm SCS} = \frac{4 \zeta}{1+\zeta^2} \overline{\Psi} \left(1+i\alpha\gamma^5\right) \overline{\upsilon^{ij}} \Upsilon \delta^3 (x-y)
\end{equation}
as well as
\begin{equation}
    \{{\cal T}_{ij}(x),C_{{\Psi}}(y)\}|_{\rm SCS} =
    \frac{-4 \zeta}{1+\zeta^2} \Upsilon \upsilon^{ij} \left(1+i\alpha\gamma^5\right) \Psi \delta^3 (x-y)
\end{equation}
and
\begin{eqnarray}
    \{{\cal T}_{ij}(x),{\cal T}^{kl}(y)\}|_{\rm SCS}
    \!\!&=&\!\! \tensor{X}{_i_j^k^l} \delta^3(x-y)
    \label{eq:TT bracket}\\
    \!\!&&\!\!
    + \tensor{Y}{^d_i_j^k^l} \frac{\partial \delta^3(x-y)}{\partial x^d} \,,\nonumber
\end{eqnarray}
where the subscript "SCS" denotes an evaluation on the second-class constraint surface,
\begin{equation}
    \tensor{Y}{^d_i_j^k^l} = 2 \frac{1}{1+\zeta^2} v^s \tensor{\epsilon}{_{s}_{(i}^{(k}} \tensor{\epsilon}{^{l)}_{j)}^p} {\cal P}^d_p\,,
\end{equation}
and
\begin{widetext}
\begin{eqnarray}
    \tensor{X}{_i_j^k^l} \!\!&=&\!\! - \frac{1}{1+\zeta^2} \Bigg[ \left(\delta_{ij} \delta^{p(k} \delta_q^{l)}
    - \delta^{kl} \delta^p_{(i} \delta_{j)q}
    \right) \left(\tilde{\cal P}^a_p K_a^q - \tilde{\cal K}^a_p \Gamma_a^q\right)
    + \delta^{(k}_{(i} \tensor{\epsilon}{_{j)}^{l)}^r} \left(G_r^{(G)}+\partial_a\tilde{\cal K}^a_r\right)
    \\
    \!\!&&\!\!\qquad\qquad
    + \left(v^s \tensor{\epsilon}{_s_{(i}^{(k}}
    - \zeta \delta_{(i}^{(k}\right) \delta_{j)r} \tensor{\epsilon}{^{l)}^p^q} \left({\cal P}^{-1}\right)_c^r {\cal P}^d_p \partial_d {\cal P}_q^c
    + \left( v^s \tensor{\epsilon}{_s_{(i}^{(k}}
    + \zeta \delta_{(i}^{(k}\right) \delta^{l)}_{|r|} \tensor{\epsilon}{_{j)}^p^q} \left({\cal P}^{-1}\right)_c^r {\cal P}^d_p \partial_d {\cal P}_q^c
    \nonumber\\
    \!\!&&\!\!\qquad\qquad
    - \frac{1}{2} \frac{\zeta}{1+\zeta^2} \bigg(P_\Psi \left(\zeta - i \gamma^5\right) \Big[\gamma^2 \left(\gamma^0 - v_u \gamma^u\right) \left[v_{(i} \delta^{(k}_{j)} \gamma^{l)}
    - v^{(k} \delta_{(i}^{l)} \gamma_{j)}\right]
    + [\upsilon_{ij},\upsilon^{kl}] \Big] \Psi
    + c.c.\bigg)
    \nonumber\\
    \!\!&&\!\!\qquad\qquad
    - \frac{1}{4} \frac{1}{1+\zeta^2} \left( P_\Psi [\upsilon_{ij},\upsilon^{kl}] \left(\zeta - i \gamma^5\right)^2 \Psi
    + c.c.
    \right)
    \Bigg]
    ,\nonumber
\end{eqnarray}
with
\begin{eqnarray}
    [\upsilon_{ij},\upsilon^{kl}]
    \!\!&=&\!\!
    4 i \gamma^2 v_{(i} v^{(k} \tensor{S}{_{j)}^{l)}}
    + 2 \gamma^4 \left(\gamma^0 - v_u \gamma^u\right) \left[v^{(k} \gamma^{l)} v_i v_j - v_{(i} \gamma_{j)} v^k v^l\right]
    \,.
\end{eqnarray}

The second-class constraint matrix is given by
\begin{eqnarray}
    C(x,y)
    \!\!&=&\!\!
    \left(\begin{matrix}
        \{\mathfrak{K}_{ij}(x),\mathfrak{K}^{kl}(y)\}
        & \{\mathfrak{K}_{ij}(x),{\cal T}^{kl}(y)\}
        & \{\mathfrak{K}_{ij}(x),C_{\overline{\Psi}}(y)\}
        & \{\mathfrak{K}_{ij}(x),{C_\Psi}(y)\}
        \\
        \{{\cal T}_{ij}(x),\mathfrak{K}^{kl}(y)\}
        & \{{\cal T}_{ij}(x),{\cal T}^{kl}(y)\}
        & \{{\cal T}_{ij}(x),C_{\overline{\Psi}}(y)\}
        & \{{\cal T}_{ij}(x),{C_\Psi}(y)\}
        \\
        \{C_{\overline{\Psi}}(x),\mathfrak{K}^{kl}(y)\}
        & \{C_{\overline{\Psi}}(x),{\cal T}^{kl}(y)\}
        & \{C_{\overline{\Psi}}(x),C_{\overline{\Psi}}(y)\}
        & \{C_{\overline{\Psi}}(x),{C_\Psi}(y)\}
        \\
        \{{C_\Psi}(x),\mathfrak{K}^{kl}(y)\}
        & \{{C_\Psi}(x),{\cal T}^{kl}(y)\}
        & \{{C_\Psi}(x),C_{\overline{\Psi}}(y)\}
        & \{{C_\Psi}(x),{C_\Psi}(y)\}
    \end{matrix}\right) \Bigg|_{\rm SCS}
    \\
    \!\!&=&\!\!
    \left(\begin{matrix}
        0
        & - \frac{V}{1+\zeta^2}
        & 0
        & 0
        \\
        \frac{V}{1+\zeta^2}
        & X
        + Y^d \frac{\partial}{\partial x^d}
        & \frac{4 \zeta}{1+\zeta^2} \overline{\Psi} \left(1+i\alpha\gamma^5\right) \overline{\upsilon} \Upsilon
        & - \frac{4 \zeta}{1+\zeta^2} \Upsilon \upsilon \left(1+i\alpha\gamma^5\right) \Psi
        \\
        0
        & \frac{-4 \zeta}{1+\zeta^2} \overline{\Psi} \left(1+i\alpha\gamma^5\right) \overline{\upsilon} \Upsilon
        & 0
        & \Upsilon
        \\
        0
        & \frac{4 \zeta}{1+\zeta^2} \Upsilon \upsilon \left(1+i\alpha\gamma^5\right) \Psi
        & \Upsilon
        & 0
    \end{matrix}\right) \delta^3(x-y)
    \,,\nonumber
\end{eqnarray}
and its inverse by
\begin{equation}
    C^{-1}(x,y)
    =
    \left(\begin{matrix}
        H(x,y)
        & (1+\zeta^2) (V^{-1}) \delta^3(x-y)
        & D_{\Psi} \delta^3(x-y)
        & D_{\overline{\Psi}} \delta^3(x-y)
        \\
        - (1+\zeta^2) (V^{-1}) \delta^3(x-y)
        & 0
        & 0
        & 0
        \\
        D_{{\Psi}} \delta^3(x-y)
        & 0
        & 0
        & \Upsilon^{-1} \delta^3(x-y)
        \\
        D_{\overline{\Psi}} \delta^3(x-y)
        & 0
        & \Upsilon^{-1} \delta^3(x-y)
        & 0
    \end{matrix}\right)
    \,,
\end{equation}
with
\begin{equation}
    D_{\Psi} = 4 \zeta (V^{-1}) \upsilon \left(1+i\alpha\gamma^5\right) \Psi
    \quad,\quad
    D_{\overline{\Psi}} = - 4 \zeta \overline{\Psi} \left(1+i\alpha\gamma^5\right) \overline{\upsilon} (V^{-1})
    \,,
\end{equation}
\begin{equation}
    H (x,y) = \left(1+\zeta^2\right)^2 \left[
    (V^{-1}) X (V^{-1})
    - (V^{-1}) \frac{\partial Y^d}{\partial x^d} (V^{-1}) \right] \delta^3(x-y)
    + \left(1+\zeta^2\right)^2 (V^{-1}) Y^d (V^{-1}) \frac{\partial \delta^3(x-y)}{\partial x^d}\,,
\end{equation}
where the expressions are ordered according to the contraction of the corresponding internal indices, which have been suppressed for brevity, such that
\begin{eqnarray}
    \int{\rm d}^3z\; C^{-1}(x,z) C(z,y)
    = \int{\rm d}^3z\; C(x,z) C^{-1}(z,y)
    = 
    \left(\begin{matrix}
        \delta^{(k}_i\delta^{l)}_j
        & 0 & 0 & 0
        \\
        0
        & \delta^{(k}_i\delta^{l)}_j & 0 & 0
        \\
        0 & 0 & \delta^{\dot{A}}_{\dot{B}} & 0
        \\
        0 & 0 & 0 & \delta^{\dot{A}}_{\dot{B}}
    \end{matrix}\right) \delta^3(x-y)\,.
\end{eqnarray}

The Dirac bracket, for any phase-space functions ${\cal O}$ and ${\cal U}$, is then given by
\begin{equation}
    \{{\cal O},{\cal U}\}_{\rm D} = \{{\cal O},{\cal U}\}
    - \{{\cal O},{\cal U}\}_{\rm C}|_{\rm SCS}
    \,,
\end{equation}
with correction bracket
\begin{eqnarray}
    \{{\cal O},{\cal U}\}_{\rm C} = \int{\rm d}^3z_1{\rm d}^3z_2
    \left(\begin{matrix}
        \{{\cal O},\mathfrak{K}(z_1)\}\\
        \{{\cal O},{\cal T}(z_1)\}\\\{{\cal O},C_{\overline{\Psi}}(z_1)\}\\\{{\cal O},C_\Psi(z_1)\}
    \end{matrix}\right)^{\rm T}
    (C^{-1})(z_1,z_2)
    \left(\begin{matrix}
        \{\mathfrak{K}(z_2),{\cal U}\}\\\{{\cal T}(z_2),{\cal U}\}\\\{C_{\overline{\Psi}}(z_2),{\cal U}\}\\\{C_\Psi(z_2),{\cal U}\}
    \end{matrix}\right)\,.
\end{eqnarray}
\end{widetext}

\subsection{Dynamics and covariance on the second-class constraint surface}

If ${\cal U}$ is a first-class constraint, then the correction bracket $\{{\cal O},{\cal U}\}_{\rm C}$ is non-trivial on the first-class and second-class constraint surfaces only if ${\cal O}$ depends on ${\cal B}_{ij}$ or the fermionic variables.

If both ${\cal O}$ and ${\cal U}$ are first-class constraints, then the correction bracket is proportional to first-class or second-class constraints; therefore, $G_i$, $L_i$, $H_a$, and $H$ remain first-class on the second-class constraint surface when using Dirac brackets.

Using the above, the correction brackets do not contribute to the covariance conditions of the metric (\ref{eq:Spacetime covariance condition - reduced}), the tetrad (\ref{eq:Tetrad covariance condition - reduced - 1})-(\ref{eq:Tetrad covariance condition - reduced - 4}, and the fermionic variables (\ref{eq:Spinor covariance condition - reduced}), (\ref{eq:Spinor momentum covariance condition - reduced - 1}), and (\ref{eq:Spinor momentum covariance condition - reduced - 2}).
They contribute to the covariance condition of the connection (\ref{eq:Connection covariance condition - reduced}) because it depends on ${\cal B}_{ij}$.
However, this contribution is proportional to
\begin{eqnarray}\label{eq:Cov cond T - Second class}
    \frac{\partial\{{\cal T}^{ij},H[\epsilon^{\bar{0}}]\}}{\partial(\partial_c\epsilon^{\bar{0}})} \bigg|_{\rm OS}
    \!\!&=&\!\! 0
    \,,
\end{eqnarray}
which vanishes on shell for the following reason.
The tensor ${\cal T}^I_{\mu\nu}$, given by (\ref{eq:Torsion EoM}), is generally non-vanishing and, as a function of the tetrad, the connection, and the fermionic variables, it satisfies the covariance identity
\begin{equation}
    \delta_{\epsilon,\theta,\beta} {\cal T}^I_{\mu\nu} \big|_{\rm OS} = \mathcal{L}_\xi {\cal T}^I_{\mu\nu}
    + \delta^{{\rm SL}(2,\mathbb{C})} {\cal T}^I_{\mu\nu} \big|_{\rm OS}
\end{equation}
in the extended phase space (off the second-class constraint surface).
The relevant spatial components then satisfy
\begin{equation}\label{eq:Cov cond T - spatial}
    \frac{\partial\{{\cal T}_{a b}^k,H[\epsilon^{\bar{0}}]\}}{\partial(\partial_d \epsilon^{\bar{0}})} \bigg|_{\rm OS}
    = - 2 {\cal T}_{\bar{0} [a}^k \delta_{b]}^d \bigg|_{\rm OS} \,,
\end{equation}
and the right-hand side vanishes on the second-class constraint surface where ${\cal T}^I_{\mu\nu}=0$; therefore, (\ref{eq:Cov cond T - Second class}) holds too.

Finally, imposing the Dirac-fermion constraint---which can also be done without imposing the gravitational second-class constraints---the correction bracket yields
\begin{eqnarray}
    \{\Psi,{\cal H}\}_{\rm C} |_{\rm OS}\!\!&=&\!\! \Upsilon^{-1} \{C_{\Psi},{\cal H}\} |_{\rm OS}
    \\
    \!\!&=&\!\!
    \frac{i}{2} \Upsilon^{-1} \left(1-i\alpha\gamma^5\right) D_\mu \left(|e| e^\mu_I \gamma^I\right)
    \,,\nonumber
\end{eqnarray}
and hence the equation of motion becomes
\begin{eqnarray}
    \dot{\Psi} \!\!&=&\!\! \{\Psi,{\cal H}\}_{\rm D}
    \\
    \!\!&=&\!\!
    \Theta^a \partial_a \Psi
    + i \Omega \Psi
    - \frac{i}{2} \Upsilon^{-1} \left(1-i\alpha\gamma^5\right) D_\mu \left(|e| e^\mu_I \gamma^I\right)
    \,,\nonumber
\end{eqnarray}
which indeed matches the Dirac equation (\ref{eq:Dirac equation - Lagrange - phase space}).

Finally, use of the Dirac brackets preserves the Dirac observables (\ref{eq:Lorentz observable}), (\ref{eq:Gauss observable}), and (\ref{eq:Electric charge observable}) because they are independent of ${\cal B}_{ij}$ on the second-class constraint surface.

\section{Mass term}
\label{sec:Mass term}

Including the mass term of the action
\begin{equation}
    S_m = - \int{\rm d}^4 x |e| m \overline{\Psi} \Psi
\end{equation}
to the extended phase-space system is not trivial.
The candidate terms
\begin{equation}
    \overline{\Psi} \Psi
    \,\,\,\,\,,\,\,\,\,\,
    P_\Psi \overline{\Xi^{-1}} \Xi^{-1} \overline{P_\Psi}
    \,\,\,\,\,,\,\,\,\,\,
    - \frac{1}{2} \left(P_\Psi \overline{\Xi^{-1}} \Psi
    + \overline{\Psi} \Xi^{-1} \overline{P_\Psi}\right)
    \,,
\end{equation}
are all equivalent upon use of the Dirac-fermion constraints (\ref{eq:Dirac fermion constraint - 1}) and (\ref{eq:Dirac fermion constraint - 2}), but they are not otherwise: They correspond to the following contributions to the extended fermion action,
\begin{eqnarray}
    S_{m}^{\rm ext} \!\!&=&\!\! - \int{\rm d}^4 x |e| \bigg[m_\Phi \overline{\Phi} \Phi
    + m_\Psi \overline{\Psi} \Psi
    \\
    \!\!&&\!\!\qquad\qquad\quad
    + \frac{m_{\Phi\Psi}}{2} \left(\overline{\Phi} \Psi
    + \overline{\Psi} \Phi\right)\bigg]
    \,.\nonumber
\end{eqnarray}
Their contributions to the equations of motion mix $\Phi$ and $\Psi$, such that it is not entirely clear whether they should be considered interactions terms instead.

However, this ambiguity does not appear in the reduced phase-space system, where one can simply include the following mass contribution to the Hamiltonian constraint
\begin{equation}
    H_{(m)} = \sqrt{\det q} m \overline{\Psi} \Psi
    \,,
\end{equation}
and use Dirac brackets for the equations of motion:
Computing
\begin{eqnarray}
    \{\Psi,H_{(m)}[N]\} = 0
\end{eqnarray}
and
\begin{eqnarray}
    \{\Psi,H_{(m)}[N]\}_{\rm C} \!\!&=&\!\! - \Upsilon^{-1} \Psi
    \nonumber\\
    \!\!&=&\!\! i m \gamma \left(\gamma^0 - v_m \gamma^m\right)
\end{eqnarray}
we conclude that the Dirac equation indeed acquires the correct mass term when using Dirac brackets.

\section{Canonical quantization and densitized spinors}
\label{sec:Canonical quantization}

\subsection{Resolution to the inconsistency of the reality conditions}

A critique about the canonical quantization of fermions coupled to gravity, raised in \cite{Inconsistency}---and followed up in \cite{Half-densitized} and several other studies---argues that the reality condition of the gravitational variables imposed on the spinor momenta implies an inconsistency upon quantization as follows.
Adapted to the variables of our system, one first promotes the classical relation (\ref{eq:Momenta Dirac fermion}) to the operator relation
\begin{equation}\label{eq:Momentum operator fermion}
    \hat{P}_{\Psi} = - \hat{\overline{\Psi}}\,\hat{\overline{\Xi}}\,,
\end{equation}
where $\hat{\Xi}$ is the operator version of (\ref{eq:Xi}).
The quantum reality condition of the momentum (\ref{eq:Reality conditions - canonical}) then implies
\begin{equation}\label{eq:Reality condition}
    \hat{\overline{P_\Psi}} = - \hat{\Xi} \hat{\Psi}
    \,.
\end{equation}
If $f$ is a real-valued function of the connection components $K_a^i$ or $\Gamma_a^i$, and $\hat{f}$ its operator version, then the following inconsistent relation ensues:
\begin{equation}\label{eq:Fermion inconsistency - quantum}
    0 = \overline{[\hat{P}_{\Psi},\hat{f}]} = - \frac{i}{2} [\widehat{|e| e^t_I},\hat{f}] \left(1-i\alpha\gamma^5\right) \gamma^I \hat{\Psi}
    \neq 0\,.
\end{equation}

The first fallacy of this argument lies in the fact that the quantization procedure plays no role in the alleged inconsistency: In the classical context, the same calculation applies,
\begin{equation}\label{eq:Fermion inconsistency - classical}
    0 = \overline{\{P_{\Psi},f\}} = - \frac{i}{2} \{|e| e^t_I,f\} \left(1-i\alpha\gamma^5\right) \gamma^I \Psi
    \neq 0\,.
\end{equation}
Therefore, this argument is ill-posed as a critique for the quantum theory and its resolution should lie already at the classical level.

The second fallacy of this argument lies in the fact that the reality condition plays no role in the alleged inconsistency either: The inconsistency appears without use of the complex conjugation,
\begin{equation}\label{eq:Fermion inconsistency - classical - no reality}
    0 = \{P_{\Psi},f\} = \frac{i}{2} \{|e| e^t_I,f\} \overline{\Psi} \gamma^I \left(1-i\alpha\gamma^5\right)
    \neq 0\,.
\end{equation}

The third fallacy of the argument lies in the implicit assumption that the bracket used in the left-hand side, $\{P_{\Psi},f\}$, shares the same symplectic structure as the bracket in the right-hand side, $\{|e| e^t_I,f\}$: The relation (\ref{eq:Momentum operator fermion}), or its classical version (\ref{eq:Momenta Dirac fermion}), constitutes a constraint on the phase space---given by (\ref{eq:Dirac fermion constraint - 1}) and (\ref{eq:Dirac fermion constraint - 2})---whose imposition requires the use of the Dirac brackets computed in Subsection~\ref{sec:Dirac brackets}.
Indeed, the Dirac bracket that correspondingly constrains the phase space automatically satisfies
\begin{equation}
    \{C_{\overline{\Psi}},f\}_{\rm D}=0\,,
\end{equation}
and hence
\begin{equation}
    \{P_{\Psi},{\cal O}\}_{\rm D} = \{\frac{i}{2} |e| e^t_I \overline{\Psi} \gamma^I \left(1-i\alpha\gamma^5\right),{\cal O}\}_{\rm D}
\end{equation}
for any phase-space function ${\cal O}$, which should hold in the quantum theory by use of corresponding Dirac commutators.
Notice that $\{P_{\Psi},f(A,B)\}_{\rm D}\neq0$ in the reduced phase space, while $\{P_{\Psi},f(A,B)\}=0$ in the extended phase space; the failure to distinguish this difference is the central mistake in (\ref{eq:Fermion inconsistency - classical - no reality}).

\subsection{Half-densitized fermions}

We now address the alleged resolution to the inconsistency above put forward in \cite{Half-densitized}, based on the introduction of a half-densitized fermionic configuration variable
\begin{equation}
    \tilde{\Psi} = \sqrt[4]{\det q} \Psi
    \,,
\end{equation}
with half-densitized momentum
\begin{equation}\label{eq:Half-densitized momentum}
    \tilde{P}_\Psi = \frac{P_\Psi}{\sqrt[4]{\det q}}
    \,,
\end{equation}
and a redefinition of the gravitational configuration variables,
\begin{eqnarray}
    \tilde{A}_a^i \!\!&=&\!\! A_a^i + \frac{1}{4} \left(\tilde{P}_{\Psi}\tilde{\Psi}+\overline{\tilde{\Psi}}\,\overline{\tilde{P}_\Psi}\right) \frac{\partial \ln \det q}{\partial {\cal P}^a_i}
    \,,\\
    \tilde{B}_a^i \!\!&=&\!\! B_a^i + \frac{1}{4} \left(\tilde{P}_{\Psi}\tilde{\Psi}+\overline{\tilde{\Psi}}\,\overline{\tilde{P}_\Psi}\right) \frac{\partial \ln \det q}{\partial {\cal K}^a_i}
    \,,
\end{eqnarray}
such that the symplectic structure becomes
\begin{eqnarray}\label{eq:Symplectic structure half-densitized}
    \!\!&&\!\!
    \int {\rm d}^4x \left[{\cal P}^a_i \dot{A}_a^i
    + {\cal K}^a_i \dot{B}_a^i
    + P_{\Psi{\dot{A}}} \dot{\Psi}^{\dot{A}} + \dot{\overline{\Psi}}_{\dot{A}} \overline{P_\Psi}{}^{\dot{A}}\right]
    \\
    \!\!&&\!\!\qquad\qquad
    = \int {\rm d}^4x \Bigg[{\cal P}^a_i \dot{A}_a^i
    + {\cal K}^a_i \dot{B}_a^i
    + \tilde{P}_{\Psi} \dot{\tilde{\Psi}}
    + \dot{\overline{\tilde{\Psi}}}\,\overline{\tilde{P}_\Psi}
    \nonumber\\
    \!\!&&\!\!\qquad\qquad\qquad\qquad
    - \frac{1}{4} \left(\tilde{P}_{\Psi}\tilde{\Psi}+\overline{\tilde{\Psi}}\,\overline{\tilde{P}_\Psi}\right) \left(\ln \det q\right)^\bullet
    \Bigg]
    \nonumber
    \\
    \!\!&&\!\!\qquad\qquad
    = \int {\rm d}^4x \left[{\cal P}^a_i \dot{\tilde{A}}_a^i
    + {\cal K}^a_i \dot{\tilde{B}}_a^i
    + \tilde{P}_{\Psi} \dot{\tilde{\Psi}}
    + \dot{\overline{\tilde{\Psi}}}\,\overline{\tilde{P}_\Psi}\right]
    \,,\nonumber
\end{eqnarray}
where an integration by parts in the time coordinate is performed,  and boundary terms are neglected, to obtain the last line.

The previous argument is then repeated with the new variables, such that Eq.~(\ref{eq:Fermion inconsistency - classical - no reality}) is replaced for
\begin{equation}\label{eq:Fermion inconsistency - classical - half-densitized}
    0 = \{\tilde{P}_\Psi,f(\tilde{A},\tilde{B})\} = 0\,,
\end{equation}
hence resolving the inconsistency according to \cite{Half-densitized}.
However, there is no actual resolution presented by this procedure at all: The phase-space variable $\tilde{P}_\Psi$ Poisson commutes with $\tilde{A}_a^i$ and $\tilde{B}_a^i$, but the Dirac-fermion conditions were not used in (\ref{eq:Fermion inconsistency - classical - half-densitized}) as they were used to arrive at (\ref{eq:Fermion inconsistency - classical - no reality}).
Indeed, following the logic that led to (\ref{eq:Fermion inconsistency - classical - no reality}), the replacement of the relation (\ref{eq:Half-densitized momentum}) into (\ref{eq:Fermion inconsistency - classical - half-densitized}) and use of the Dirac-fermion conditions yields an inconsistency once again,
\begin{eqnarray}\label{eq:Fermion inconsistency - classical - half-densitized - no resolution}
    0 \!\!&=&\!\! \{\tilde{P}_\Psi,f(\tilde{A},\tilde{B})\}
    = \{P_\Psi/\sqrt[4]{\det q},f(\tilde{A},\tilde{B})\}
    \\
    \!\!&=&\!\!
    \frac{i}{2} \overline{\tilde{\Psi}} \{\gamma \sqrt[4]{\det q} \left(\gamma^0
    - v_i \gamma^i\right),f(\tilde{A},\tilde{B})\} \left(1-i\alpha\gamma^5\right)
    \neq 0
    \,.\nonumber
\end{eqnarray}

A basic lesson of Hamiltonian formulations instructs us that the system remains unchanged under this redefinition of variables because it is equivalent to a canonical transformation---implied by the preservation of the symplectic structure (\ref{eq:Symplectic structure half-densitized}).
In particular, the results of Poisson brackets are invariant under canonical transformations.
Were (\ref{eq:Fermion inconsistency - classical - no reality}) a true inconsistency, it should remain an inconsistency under any canonical transformation, including the use of half-densitized fermionic variables.
Therefore, the original inconsistency (\ref{eq:Fermion inconsistency - quantum}) cannot be due to the use of specific canonical variables as argued in \cite{Half-densitized}; it is due to a negligent treatment of the phase-space reduction implied by the relation (\ref{eq:Momentum operator fermion}).

\section{Discrete symmetries}
\label{sec:Discrete symmetries}

In this section, we seek to provide a complete analysis of the discrete symmetries of our system in the spirit of previous studies based on traditional Ashtekar--Barbero variables \cite{Parity,Rastgoo}.

In the Einstein--Cartan theory, the SO(1,3) internal symmetry is associated to an internal frame basis rather than a coordinate one.
Therefore, extending this group to O(1,3) requires an analysis of the discrete inversion of the internal basis that is not related to an actual inversion of the time or space coordinates---the latter may be associated instead to a discrete extension of the spacetime diffeomorphism group.
The decomposition of the internal space (\ref{eq:Internal foliation}) induces a natural notion for internal parity and time reversal, defined by the discrete transformation of reversing the internal directions of the frame.
In the following, the transformations $P$ and $T$ refer to parity and time-reversal transformations of the internal basis, respectively.
This is the appropriate treatment if $P$ and $T$ are to be applied to the fermionic variables (and to associate them to the charge conjugation transformation $C$), which involve actions on the spinorial internal indices of the larger SL(2,$\mathbb{C}$) group.

\subsection{Geometric and Lagrangian formulations}

\subsubsection{Internal parity}
\label{sec:Parity}

The tetrads $e^\mu_I$ constitute four orthonormal vectors representing a dynamical frame, which we use to define parity and time-reversal transformations from a geometric perspective in the following.

We define the parity transformation as the change in sign of the Euclidean components of the tetrad:
\begin{eqnarray}\label{eq:Parity tetrad - geometric}
    P^{-1} e^\mu_0 P = e^\mu_0
    \quad,\quad
    P^{-1} e^\mu_i P = - e^\mu_i
    \,.
\end{eqnarray}
This implies
\begin{eqnarray}\label{eq:Parity Tetrad components}
    P^{-1} N P = N
    &,&
    P^{-1} v_i P = - v_i\,,
    \\
    P^{-1} N^a P = N^a
    &,&
    P^{-1} \varepsilon^a_i P = - \varepsilon^a_i
    \,.\nonumber
\end{eqnarray}

Moreover, the parity transformation of spinors can be inherited from our understanding of flat spacetime,
\begin{equation}\label{eq:Parity fermion - geom}
    P^{-1} \Psi P = i \beta \Psi\,,
\end{equation}
or
\begin{equation}
    \psi_{\dot{a}}\to i\chi^{\ddot{a}}
    \quad,\quad
    \chi^{\ddot{a}}\to i\psi_{\dot{a}}
    \,.
\end{equation}

Notice that the parity transformations (\ref{eq:Parity tetrad - geometric}) and (\ref{eq:Parity fermion - geom}) do not imply the transformation of the connection.

An inspection of the action shows that it is invariant to the parity transformation if it is supplemented with the transformation of the connection
\begin{eqnarray}\label{eq:Parity connection - geometric action}
    P^{-1} \tensor{\omega}{_\mu^0^i} P = - \tensor{\omega}{_\mu^0^i}
    \quad,\quad
    P^{-1} \tensor{\omega}{_\mu^i^j} P = \tensor{\omega}{_\mu^i^j}
    \,,
\end{eqnarray}
and the parameter sign changes
\begin{equation}\label{eq:Parity Z}
    \zeta \to - \zeta
    \quad,\quad \alpha \to - \alpha
    \,.
\end{equation}
However, $\zeta$ and $\alpha$ are constants rather than field variables; therefore, (\ref{eq:Parity Z}) cannot be part of a $P$ transformation and hence
they correspond to a parity-violating terms.

\subsubsection{Internal time reversal}

We define the time-reversal transformation as the change in sign of the normal internal components of the tetrad:
\begin{equation}\label{eq:Time-reversal grav}
    T^{-1} e^\mu_0 T = - e^\mu_0
    \quad,\quad
    T^{-1} e^\mu_i T = e^\mu_i
    \,.
\end{equation}
This implies
\begin{eqnarray}
    T^{-1} N T = - N
    &,&
    T^{-1} v_i T = - v_i\,,
    \\
    T^{-1} N^a T = N^a
    &,&
    T^{-1} \varepsilon^a_i T = \varepsilon^a_i
    \,.
\end{eqnarray}
Similarly, the transformation of the fermionic field is given by
\begin{equation}
    T^{-1} \Psi T = {\cal T} \gamma^5 \Psi
    \,,
\end{equation}
where
\begin{equation}
    {\cal T} = \left(\begin{matrix}
        - \varepsilon^{\dot{a}\dot{b}} & 0 \\
        0 & - \varepsilon_{\ddot{a}\ddot{b}}
    \end{matrix}\right)\,.
\end{equation}
or
\begin{equation}
    \psi_{\dot{a}} \to \psi^{\dot{a}}
    \quad,\quad
    \chi^{\ddot{a}} \to - \chi_{\ddot{a}}\,.
\end{equation}
The $T$ transformation is anti-Hermitian: $T^{-1} F T = F^*$ for any complex number $F$.

An inspection of the action shows that it is invariant to the time-reversal transformation if it is supplemented with the transformation of the connection
\begin{equation}\label{eq:Time-reversal connection - geometric action}
    T^{-1} \tensor{\omega}{_\mu^0^i} T = - \tensor{\omega}{_\mu^0^i}
    \quad,\quad
    T^{-1} \tensor{\omega}{_\mu^i^j} T = \tensor{\omega}{_\mu^i^j}
    \,,
\end{equation}
and the parameter sign changes (\ref{eq:Parity Z}).
Therefore, $\zeta$ and $\alpha$ are also time-reversal symmetry breaking terms.

The action is $PT$ invariant without relying on the supplemental transformation of the parameters (\ref{eq:Parity Z}).

\subsubsection{Charge conjugation}

The charge conjugation transformation is given by
\begin{equation}
    C^{-1} \Psi C = {\cal C} \overline{\Psi}^{\rm T} = {\cal C} \beta^{\rm T} \Psi^*\,,
\end{equation}
where
\begin{equation}\label{eq:Charge conjugation matrix}
    {\cal C} = \left(\begin{matrix}
        \varepsilon_{\dot{a}\dot{b}} & 0 \\
        0 & \varepsilon^{\ddot{a}\ddot{b}}
    \end{matrix}\right)\,,
\end{equation}
or
\begin{equation}
    C^{-1} \psi_{\dot{a}} C = \chi_{\dot{a}}^\dagger
    \quad,\quad
    C^{-1} \chi^{\ddot{a}} C = \psi^{\dagger\ddot{a}}\,.
\end{equation}
And the gravitational variables remain invariant under charge conjugation.

The action (\ref{eq:Dirac action}) is $C$ invariant and hence $CPT$ invariant.

\subsection{Hamiltonian formulation}

The geometric formulation provides a guide for the discrete transformations of the tetrad and the spinor field.
However, it does not automatically  prescribe the transformations of all phase-space variables of the canonical system.
Here, we propose a simple and unambiguous method to define the latter: We perform the known discrete transformations of $e^\mu_I$ and $\Psi$, which are more directly connected to the reversal of the internal frame, then perform the canonical decomposition of the transformed action; the decomposition of the transformed theory then defines the transformation of their canonical conjugates (the connection and fermion momenta) via the preservation of the symplectic structure.

The relation between the phase space and geometry holds only on-shell, but it is a universal relation that does not depend on the details of the dynamics \cite{EMGCov}.
Therefore, the geometric parity transformation of the tetrad components (\ref{eq:Parity Tetrad components}) induces a direct transformation of the following phase-space variables,
\begin{eqnarray}\label{eq:Parity phase space 1}
    P^{-1} {\cal P}^a_i P = - {\cal P}^a_i
    &,&
    P^{-1} {\cal K}^a_i P = {\cal K}^a_i
    \,,\nonumber\\
    P^{-1} \Psi P = i \beta \Psi&,&
    P^{-1} \overline{\Psi} P = - i \overline{\Psi} \beta\,,
\end{eqnarray}
where we also included the fermion transformation (\ref{eq:Parity fermion - geom}).

The preservation of the symplectic term,
\begin{eqnarray}
    \!\!&&\!\!
    P^{-1} \int{\rm d}^4x \left[{\cal P}^a_i A^a_i+{\cal K}^a_i B^a_i+P_\Psi \Psi+\overline{\Psi} \overline{P_\Psi}\right] P
    \\
    \!\!&&\!\!\qquad\qquad
    = \int{\rm d}^4x \big[- {\cal P}^a_i P^{-1} A^a_i P+{\cal K}^a_i P^{-1} B^a_i P
    \nonumber\\
    \!\!&&\!\!\qquad\qquad\qquad\qquad
    +iP^{-1} P_\Psi P \beta \Psi
    - i \overline{\Psi} \beta P^{-1} \overline{P_\Psi} P\big]
    \,,\nonumber
\end{eqnarray}
determines
\begin{eqnarray}\label{eq:Parity phase space 2}
    P^{-1} A^a_i P = - A_a^i
    &,& P^{-1} B^a_i P = B_a^i
    \,,\nonumber\\
    P^{-1} P_\Psi P = - i P_\Psi \beta
    &,&
    P^{-1} \overline{P_\Psi} P = i \beta \overline{P_\Psi}\,.
\end{eqnarray}
Similarly, we obtain the internal-time-reversal transformations
\begin{eqnarray}\label{eq:Time-reversal phase space 1}
    T^{-1} {\cal P}^a_i T = - {\cal P}^a_i
    &,&
    T^{-1} {\cal K}^a_i T = {\cal K}^a_i
    \,,\nonumber\\
    T^{-1} \Psi T = {\cal T} \gamma^5 \Psi&,&
    T^{-1} \overline{\Psi} T = \overline{\Psi} \gamma^5 {\cal T}^{-1}\,.
\end{eqnarray}
and
\begin{eqnarray}\label{eq:Time-reversal phase space 2}
    T^{-1} A^a_i T = - A_a^i
    &,& T^{-1} B^a_i T = B_a^i
    \,,\nonumber\\
    T^{-1} P_\Psi T = P_\Psi \gamma^5 {\cal T}^{-1}
    &,&
    T^{-1} \overline{P_\Psi} T = {\cal T} \gamma^5 \overline{P_\Psi}\,.
\end{eqnarray}
The charge-conjugation transformations are given by
\begin{eqnarray}\label{eq:Charge conjugation phase space}
    C^{-1} \Psi C = {\cal C} \overline{\Psi}^{\rm T}&,&
    C^{-1} \overline{\Psi} C = \Psi^{\rm T} {\cal C}\,,\\
    C^{-1} P_\Psi C = \overline{P_\Psi}^{\rm T} {\cal C} &,&
    C^{-1} \overline{\Psi} C = {\cal C} P_\Psi^{\rm T}\,.
\end{eqnarray}

Notice that the parity and time-reversal transformations (\ref{eq:Parity phase space 2}) and (\ref{eq:Time-reversal phase space 2}) of the gravitational configuration variables do not match the connection transformations (\ref{eq:Parity connection - geometric action}) and (\ref{eq:Time-reversal connection - geometric action}).
The reason is that they are simply not equivalent transformations:
Given the transformations of the tetrad (\ref{eq:Parity tetrad - geometric}) and (\ref{eq:Time-reversal grav}), equations (\ref{eq:Parity phase space 2}) and (\ref{eq:Time-reversal phase space 2}) determine what the configuration variables associated to the new momenta are, irrespective of the values of $\zeta$ and $\alpha$; on the other hand, equations (\ref{eq:Parity connection - geometric action}) and (\ref{eq:Time-reversal connection - geometric action}) are defined as the necessary transformations of the connection components for the action to remain invariant in the absence of $\zeta$ and $\alpha$.

Using the transformations that we have derived, we find that the first-class constraints are $C$ and $PT$ invariant for arbitrary values of $\zeta$, while they are $P$ and $T$ invariant only if $\zeta=0$.
(Recall that the Hamiltonian constraint is smeared by $|N|$ rather than $N$ as explained after Eq.~(\ref{eq:Action canonical 1}). Therefore, $H$ does not receive an overall negative sign under an internal time reversal transformation despite the lapse transformation $T^{-1}N T=-N$ implied by (\ref{eq:Parity Tetrad components}).)

Similarly, the gravitational second-class constraints (\ref{eq:Second-class constraint K=0}) and (\ref{eq:Torsion-spatial-sym}) are $PT$ invariant for arbitrary values $\zeta$.
If $\zeta=0$, they are $P$ and $T$ invariant up to an overall sign; however, the result of the Dirac brackets remains unchanged under this overall sign change and hence their dynamical implications are indeed invariant under $P$ and $T$ transformations.

The non-minimal coupling $\alpha$ appears only in the Dirac-fermion constraints (\ref{eq:Dirac fermion constraint - 1}) and (\ref{eq:Dirac fermion constraint - 2}).
The discrete transformations of these constraints are given by
\begin{equation}\label{eq:Dirac fermion constraint - P}
    P^{-1} C_{\overline{\Psi}} P = - i \left[ P_\Psi - \frac{i}{2} |e| e^t_I \overline{\Psi} \gamma^I \left(1+i\alpha\gamma^5\right) \right] \beta
    \,,
\end{equation}
\begin{equation}\label{eq:Dirac fermion constraint - T}
    T^{-1} C_{\overline{\Psi}} T = \left[P_\Psi + \frac{i}{2} |e| e^t_I \overline{\Psi} \gamma^I \left(1+i\alpha\gamma^5\right)
    \right] \gamma^5 {\cal T}^{-1}
    \,,
\end{equation}
and
\begin{equation}\label{eq:Dirac fermion constraint - C}
    C^{-1} C_{\overline{\Psi}} C = C_\Psi^{\rm T} {\cal C}
    \,,
\end{equation}
and their respective complex conjugates.
Notice that
\begin{equation}\label{eq:Dirac fermion constraint - PT}
    T^{-1} P^{-1} C_{\overline{\Psi}} P T = i C_{\overline{\Psi}} \gamma^5 {\cal T}^{-1} \beta
    \,.
\end{equation}
We conclude that the Dirac-fermions constraints are similarly $PT$ and $C$ invariant up to overall spinorial matrices that do not change the dynamics.
If $\alpha=0$, the Dirac-fermion constraints are $P$ and $T$ invariant in the same sense.

Therefore, the $PT$ and $C$ transformations remain discrete symmetries of the system even for non-zero $\zeta$ and $\alpha$.

\section{Emergent formulation}
\label{sec:EMGFT}

Recent developments have made it possible to circumvent the uniqueness theorems of general relativity \cite{hojman1976geometrodynamics,kuchar1974geometrodynamics} and formulate modified gravity (in metric variables) and Yang--Mills theories without additional degrees of freedom by introducing the concept of emergent fields \cite{EMGCov,EMGFT}, which crucially depends on features of the canonical formulation.
Explicit expressions of the modified constraints and emergent fields in this context have been obtained in several symmetry-reduced systems, including spherical and Gowdy symmetric systems, as well as cosmological ones, which all generate nonsingular dynamical solutions \cite{EMGCov,EMGGowdy,EMGCosmoK}.
Furthermore, the coupling of different forms of matter in the emergent formulation has been successfully realized, including scalar matter and perfect fluids \cite{EMGScalar,EMGPF}, with fermions being the only type of matter of the standard model yet be coupled.
Because an Einstein--Yang--Mills SO(1,3) system has the same constraint algebra of the system studied here in the extended phase space, we can import all the conceptual machinery of the emergent formulation of \cite{EMGFT} to outline how the coupling of fermions would take place.

The generic meaning of a modified theory is to have modified dynamics.
Since the dynamics is governed by the Hamiltonian constraint, this implies that we must consider a modified $\tilde{H}$ that is different from the classical Hamiltonian constraint $H$, which can generally be recovered as a limit of the modification parameters introduced in $\tilde{H}$.
However, because the Hamiltonian constraint generates not only dynamics but also gauge transformations, the allowed modifications must be restricted by the requirement that the modified system remains covariant.
This is achieved by implementing the following four-step procedure.

First, we require that $\tilde{H}$, together with the vector and Lorentz--Gauss constraints, preserves an anomaly-free algebra of the form (\ref{eq:HaHa-Poisson})-(\ref{eq:LL-Poisson}) in the extended phase space with $H$ replaced by $\tilde{H}$ and possibly modified structure functions $\tilde{q}^{ab}$, $\tilde{F}^{0i}_{\bar{0}a}$, and $\tilde{\cal F}^i_{\bar{0}a}$---such that the dependence of these structure functions on the phase-space may differ from the classical expressions, i.e., the relation (\ref{eq:Inverse spatial metric - canonical}) might suffer modifications, while (\ref{eq:calFi}) is replaced by $\tilde{F}_{\bar{0}a}^{0i} = \frac{1}{N} \{K_a^i,\tilde{H}[N]\}$ and $\tilde{\cal F}_{\bar{0}a}^i = \frac{1}{N} \{\Gamma_a^i,\tilde{H}[N]\}$.

Second, these modified structure functions are used to define an (emergent) spacetime metric $\tilde{g}_{\mu\nu}$ of the form (\ref{eq:ADM line element}) with $q_{ab}$ replaced by the emergent spatial metric $\tilde{q}_{ab}$, which is given by the inverse of $\tilde{q}^{ab}$.
Similarly, the emergent force field $\tilde{F}^{IJ}_{\mu\nu}$ is identical to the classical $F^{IJ}_{\mu\nu}$ up to the $\tilde{F}_{\bar{0}a}^{0i}$ and $\tilde{F}_{\bar{0}a}^{ij}$ components, which take the values of the corresponding structure functions.
The emergent spacetime metric defines emergent tetrads $\tilde{e}^\mu_I$ up to Lorentz transformations, which, in turn, can be used to define the emergent torsion $\tilde{T}^I_{\mu\nu}=D_{[\mu}\tilde{e}^I_{\nu]}$ and all other geometric objects.

Third, the covariance conditions must be imposed on the physical fields using the modified constraint $\tilde{H}$ instead of the classical $H$; in this case, these are given by (\ref{eq:Connection covariance condition}) for the connection and (\ref{eq:Tetrad covariance condition}), with $e_\mu^I$ replaced by $\tilde{e}^\mu_I$, for the emergent tetrad, as well as (\ref{eq:Spinor covariance condition}) and (\ref{eq:Spinor momentum covariance condition}) for $\Psi$ and $P_\Psi$ if fermionic matter is present.
The rest of the geometric quantities are constructed from these fields and are therefore covariant if the aforementioned conditions are successfully realized.

(An additional U($n$) Gauss constraint must be introduced in the constraint algebra in the presence of each new Yang--Mills field when applying the first step and a corresponding modified strength tensor field would be derived in the second step. The third step would then have to include the evaluation of a new covariance condition for each of these force fields as well as for each new scalar or perfect fluid matter field.)

Fourth, the theory must allow the implementation of the second-class constraints with possible modifications restricted by---or that may be necessary for---the preservation of covariance in the reduced phase space via Dirac brackets.
This is a highly nontrivial new step compared to previous examples of emergent modifications, which were based on a system with metric variables and without fermions and hence required no inclusion of second-class constraints.

A successful realization of this method will complete the emergent field theory of gravity with all known forms of matter and force fields.

\section{Conclusions}
\label{sec:Conclusions}

We have generalized the results of the canonical analysis in \cite{Tetrads} by coupling fermions to gravity based on the Hilbert--Palatini action with the Barbero--Immirzi, cosmological, and non-minimal fermion coupling constants.
Unlike the Ashtekar--Barbero procedure, our analysis does not rely on any gauge fixing and hence it retains the full gauge content of the Einstein--Cartan theory, given by its underlying diffeomorphism and SL(2,$\mathbb{C}$) covariance.

Moreover, we have identified a set of hitherto overlooked (second-class) constraints that are necessary to describe a theory of Dirac fermions.
A proper treatment of the phase-space reduction implied by these Dirac-fermion constraints has clarified erroneous beliefs about an inherent inconsistency of the reality conditions in the quantum theory of fermions coupled to gravity that was put forward in \cite{Inconsistency,Half-densitized}; such an inconsistency disappears when the phase-space reduction is properly taken into account.
We have also provided a detailed analysis of the discrete symmetries of the system and have found that the theory remains $C$ and $PT$ invariant even in the presence of the Barbero--Immirzi and non-minimal coupling constants, generalizing previous results based on Ashtekar--Barbero variables \cite{Parity,Rastgoo}.

We expect that our systematic study will serve as a firm basis for canonical approaches to modified and quantum theories of gravity with fermionic matter preserving diffeomorphism and Lorentz covariance.
We have presented one example of such an application in the context of the emergent formulation \cite{EMGCov,EMGFT}.

\section*{Acknowledgements}
The author thanks Martin Bojowald, Idrus Husin Belfaqih, and Manuel D\'iaz for useful discussions and going through a draft version of this paper.
This work was supported in part by NSF grant PHY-2206591.

\appendix

\section{Brackets}
\label{app:Brackets}

\subsection{Hamiltonian constraint}
\label{app:Hamiltonian constraint}

Here, we gather relevant transformation generated by $H^{(2)}$.
Using
\begin{widetext}
\begin{eqnarray}
    \{{\cal P}^a_i,H^{(2)}[N]\} \!\!&=&\!\! 2 \frac{8\pi G N}{\sqrt{\det q}} \frac{{\cal P}^a_{[i} {\cal P}^d_{q]}}{\gamma^2} K^q_d
    + \frac{N}{1+\zeta^2} \frac{\gamma \sqrt{(\det {\cal P})^{-1}}}{\sqrt{8\pi G}} {\cal P}^a_j \Big[i P_\Psi \left(\gamma^0-v_m\gamma^m\right) \left( \gamma^j S_{0i} - \zeta \gamma^j S_i\right) \Psi
    + c.c
    \Big]
    \,,
\end{eqnarray}
and
\begin{eqnarray}
    \{{\cal K}^a_i,H^{(2)}[N]\} \!\!&=&\!\! - \partial_c \left[\frac{8\pi G N}{\sqrt{\det q}} \frac{{\cal P}^c_p {\cal P}^a_q}{\gamma^2} \tensor{\epsilon}{^p^q_i}\right]
    - 2 \frac{8\pi G N}{\sqrt{\det q}} \frac{{\cal P}^a_{[i} {\cal P}^d_{q]}}{\gamma^2} \Gamma_d^q
    \nonumber\\
    \!\!&&\!\!
    + \frac{N}{1+\zeta^2} \frac{\gamma \sqrt{(\det {\cal P})^{-1}}}{\sqrt{8\pi G}} {\cal P}^a_j \left[i P_\Psi \left(\gamma^0-v_m\gamma^m\right) \left( \zeta \gamma^j S_{0i} + \gamma^j S_i\right) \Psi
    + c.c
    \right]
    \,,
\end{eqnarray}
we obtain
\begin{eqnarray}
    \{v_i,H^{(2)}[N]\} \!\!&=&\!\! - \frac{1}{2} \tensor{\epsilon}{_i_m^n} \{\left({\cal P}^{-1}\right)^m_a {\cal K}^a_n,H^{(2)}[N]\}
    \nonumber\\
    \!\!&=&\!\! - \frac{1}{2} \tensor{\epsilon}{_i_m^n} \left[\left({\cal P}^{-1}\right)^m_a \{{\cal K}^a_n,H^{(2)}[N]\}
    - \left({\cal P}^{-1}\right)^p_a \left({\cal P}^{-1}\right)^m_b \{{\cal P}^b_p,H^{(2)}[N]\} {\cal K}^a_n\right]
    \nonumber\\
    \!\!&=&\!\! \frac{8\pi G}{\gamma\sqrt{\det q}} \partial_c \left[\frac{N}{\gamma} {\cal P}^c_i\right]
    - \frac{8\pi G N}{\gamma^2\sqrt{\det q}} {\cal P}^d_q K^q_d v_i
    + \frac{8\pi G N}{\gamma^2\sqrt{\det q}} {\cal P}^d_m \Gamma_d^n \tensor{\epsilon}{^m_n_i}
    \nonumber\\
    \!\!&&\!\!
    + \frac{1}{2} \frac{8\pi G N}{\gamma^2\sqrt{\det q}} \frac{L_i - \zeta G_i}{1+\zeta^2}
    \nonumber\\
    \!\!&&\!\!
    + \frac{3}{4} \frac{8\pi G N}{\sqrt{\det q}} \frac{1}{1+\zeta^2} \left[P_\Psi \left(\gamma^0-v_s\gamma^s\right) \left(\gamma_i-v_i \gamma^0\right) \left(1 + \zeta i \gamma^5\right) \Psi
    + c.c
    \right]
    \nonumber\\
    \!\!&&\!\!
    + \frac{1}{2} \frac{8\pi G N}{\sqrt{\det q}} \frac{1}{1+\zeta^2} \delta_{i[p} \mathfrak{K}^p_{l]} \left[P_\Psi \left(\gamma^0-v_s\gamma^s\right) \gamma^l \left(i \gamma^5 - \zeta\right) \Psi
    + c.c\right]\,,
\end{eqnarray}
\begin{eqnarray}
    \{\sqrt{|\det {\cal P}|},H^{(2)}[N]\}|_{\rm OS} \!\!&=&\!\! \frac{\sqrt{|\det {\cal P}|}}{2} ({\cal P}^{-1})_a^i \{{\cal P}^a_i,H^{(2)}[N]\}
    \nonumber\\
    \!\!&=&\!\!
    \frac{1}{\sqrt{8\pi G}} \frac{N}{\gamma} {\cal P}^d_n K^n_d
    + \frac{N}{1+\zeta^2} \frac{\gamma}{\sqrt{8\pi G}} \frac{1}{4} \Big[i P_\Psi \left(\gamma^0-v_m\gamma^m\right) \left( [\gamma^i,S_{0i}] - \zeta [\gamma^i,S_i]_+\right) \Psi
    + c.c
    \Big]
    \nonumber\\
    \!\!&=&\!\!
    \frac{1}{\sqrt{8\pi G}} \frac{N}{\gamma} {\cal P}^d_n K^n_d
    + \frac{3}{4} \frac{\gamma N}{\sqrt{8\pi G}} \frac{1}{1+\zeta^2} \Big[P_\Psi \left(\gamma^0-v_m\gamma^m\right) \gamma^0 \left(1 + \zeta i \gamma^5\right) \Psi
    + c.c
    \Big]\,,
\end{eqnarray}
and
\begin{eqnarray}\label{eq:Upsilon bracket H2}
    \frac{2}{i} \{\Xi,H^{(2)}[N]\} |_{\rm OS} \!\!&=&\!\! \left(1-i\alpha\gamma^5\right) \bigg[(8\pi G)^{3/2} \{\sqrt{|\det {\cal P}|},H^{(2)}[N]\} \left(\gamma^0
    - v_i \gamma^i\right)
    - \gamma \sqrt{\det q} \{v_i,H^{(2)}[N]\} \gamma^i\bigg]
    \\
    \!\!&=&\!\!
    \frac{8\pi G N}{\gamma} {\cal P}^d_n K^n_d \left(1-i\alpha\gamma^5\right) \gamma^0
    - \left[8\pi G \partial_c \left(\frac{N}{\gamma} {\cal P}^c_i\right)
    + \frac{8\pi G N}{\gamma} {\cal P}^d_m \Gamma_d^n \tensor{\epsilon}{^m_n_i}\right] \left(1-i\alpha\gamma^5\right) \gamma^i
    \nonumber\\
    \!\!&&\!\!
    + (8\pi G)^{3/2} \frac{3}{4} \frac{\gamma N}{\sqrt{8\pi G}} \frac{1}{1+\zeta^2} \Big[P_\Psi \left(\gamma^0-v_m\gamma^m\right) \gamma^0 \left(1 + \zeta i \gamma^5\right) \Psi
    + c.c
    \Big] \left(1-i\alpha\gamma^5\right) \left(\gamma^0
    - v_i \gamma^i\right)
    \nonumber\\
    \!\!&&\!\!
    - \gamma \sqrt{\det q} \Bigg[\frac{3}{4} \frac{8\pi G N}{\sqrt{\det q}} \frac{1}{1+\zeta^2} \left[P_\Psi \left(\gamma^0-v_s\gamma^s\right) \left(\gamma_i-v_i \gamma^0\right) \left(1 + \zeta i \gamma^5\right) \Psi
    + c.c
    \right]
    \nonumber\\
    \!\!&&\!\!\qquad\qquad\quad
    + \frac{1}{2} \frac{8\pi G N}{\sqrt{\det q}} \frac{1}{1+\zeta^2} \delta_{i[p} \mathfrak{K}^p_{l]} \left[P_\Psi \left(\gamma^0-v_s\gamma^s\right) \gamma^l \left(i \gamma^5 - \zeta\right) \Psi
    + c.c\right]\Bigg] \left(1-i\alpha\gamma^5\right) \gamma^i
    \,.\nonumber
\end{eqnarray}

Using the above, Eq.~(\ref{eq:C_Psi,H}) follows.

\subsection{Second-class constraints}

Using
\begin{equation}
    \frac{\partial \upsilon^{ij}}{\partial v^k} = \gamma^2 \left(\gamma^0 - v_u \gamma^u\right) \left[2 \gamma^2 v_k v^{(i}
    + \delta^{(i}_k
    - \gamma^2 \left(\gamma^0 - v_n \gamma^n\right) \gamma_k v^{(i}\right] \gamma^{j)}
    \,,\nonumber
\end{equation}
we obtain
\begin{eqnarray}
    \frac{\partial \upsilon^{ij}}{\partial {\cal P}^a_r} \!\!&=&\!\! \frac{\partial v^m}{\partial {\cal P}^a_r} \frac{\partial \upsilon^{ij}}{\partial v^m}
    = \gamma^2 \left(\gamma^0 - v_u \gamma^u\right) v_s \left({\cal P}^{-1}\right)^{[s}_a \delta^{r]m} \left[2 \gamma^2 v_m v^{(i}
    + \delta^{(i}_m
    - \gamma^2 \left(\gamma^0 - v_n \gamma^n\right) \gamma_m v^{(i}\right] \gamma^{j)}
    \,,\\
    \frac{\partial \upsilon^{ij}}{\partial {\cal K}^a_r} \!\!&=&\!\! \frac{\partial v^m}{\partial {\cal K}^a_r} \frac{\partial \upsilon^{ij}}{\partial v^m}
    = - \gamma^2 \left(\gamma^0 - v_u \gamma^u\right) \frac{1}{2} \tensor{\epsilon}{^r^m_s} \left({\cal P}^{-1}\right)^s_a \left[2 \gamma^2 v_m v^{(i}
    + \delta^{(i}_m
    - \gamma^2 \left(\gamma^0 - v_n \gamma^n\right) \gamma_m v^{(i}\right] \gamma^{j)}
    \,,
\end{eqnarray}
\begin{equation}
    \left[
    - \delta_{r(i} \tilde{\cal K}^a_{j)}
    - \zeta \delta_{ij} {\cal P}^a_r \right] \frac{\partial \upsilon^{kl}}{\partial {\cal P}^a_r} = \frac{\zeta}{2} \gamma^2 \left(\gamma^0 - v_u \gamma^u\right) v_s \left[
    \delta^m_{(i} \delta^{s}_{j)}
    + \delta_{ij} \delta^{sm} \right] \left[2 \gamma^2 v_m v^{(k}
    + \delta^{(k}_m
    - \gamma^2 \left(\gamma^0 - v_n \gamma^n\right) \gamma_m v^{(k}\right] \gamma^{l)}
    \nonumber
\end{equation}
\begin{equation}
    \left[ \delta_{ij} {\cal P}^a_r - \delta_{r(i} \tilde{\cal P}^a_{j)} \right] \frac{\partial \upsilon^{kl}}{\partial {\cal K}^a_r} = \frac{\zeta}{2} \gamma^2 \left(\gamma^0 - v_u \gamma^u\right) v_s \left[\delta^s_{(i} \delta^m_{j)} - \delta_{ij} \delta^{sm}\right] \left[2 \gamma^2 v_m v^{(k}
    + \delta^{(k}_m
    - \gamma^2 \left(\gamma^0 - v_n \gamma^n\right) \gamma_m v^{(k}\right] \gamma^{l)}
    \,.
\end{equation}
It follows that
\begin{eqnarray}
    \frac{\delta {\cal T}^{ij}(x)}{\delta {\cal P}^a_r(z)} \!\!&=&\!\! \Bigg[\left(\delta^{ij} \delta^r_q - \delta^{r(i} \delta_q^{j)} \right) \Gamma_a^q
    + \delta_m^{(i} \tensor{\epsilon}{^{j)}^r^q} \left({\cal P}^{-1}\right)_c^m \partial_a {\cal P}^c_q
    - \left({\cal P}^{-1}\right)_a^m \left({\cal P}^{-1}\right)_c^r \delta_m^{(i} \tensor{\epsilon}{^{j)}^p^q} {\cal P}^d_p \partial_d {\cal P}_q^c
    \nonumber\\
    \!\!&&\!\!
    + \frac{1}{2} \frac{1}{1+\zeta^2} \left( P_\Psi \frac{\partial \upsilon^{ij}}{\partial {\cal P}^a_r} \Psi
    + c.c. \right) \Bigg] \delta^3(x-z)
    + \left({\cal P}^{-1}\right)_a^{(i} \tensor{\epsilon}{^{j)}^p^r} {\cal P}^d_p \frac{\partial \delta^3(x-z)}{\partial x^d}\,,
\end{eqnarray}
\end{widetext}
\begin{equation}
    \frac{\partial {\cal T}^{ij}(x)}{\partial {\cal K}^a_r(z)} = \frac{1}{2} \frac{1}{1+\zeta^2} \left(P_\Psi \frac{\partial \upsilon^{ij}}{\partial {\cal K}^a_r} \Psi
    + c.c. \right)
    - \delta^{r(i} \delta_q^{j)} K_a^q
    \,,
\end{equation}
\begin{equation}
    \frac{\partial v_q}{\partial {\cal P}^a_r} = \delta^{[r}_q \left({\cal P}^{-1}\right)^{u]}_a v_u
    \,,
\end{equation}
\begin{equation}
    \frac{\partial v_q}{\partial {\cal K}^a_r} = - \frac{1}{2} \tensor{\epsilon}{_q_u^r} \left({\cal P}^{-1}\right)^u_a
    \,,
\end{equation}
\begin{equation}
    \frac{\partial {\cal T}_{ij}(x)}{\partial B^r_a(z)} = \frac{1}{1+\zeta^2} \left[ \delta_{ij} {\cal P}^a_r - \delta_{r(i} \tilde{\cal P}^a_{j)} \right]
    \,,
\end{equation}
\begin{equation}
    \frac{\partial {\cal T}_{ij}(x)}{\partial A^r_a(z)} = \frac{1}{1+\zeta^2} \left[
    - \delta_{r(i} \tilde{\cal K}^a_{j)}
    - \zeta \delta_{ij} {\cal P}^a_r \right]
    \,,
\end{equation}
\begin{equation}
    \frac{\partial {\cal T}_{ij}(x)}{\partial K^r_a(z)} = \frac{1}{1+\zeta^2} \left[
    - \delta_{r(i} \tilde{\cal K}^a_{j)}
    - \zeta \delta_{r(i} \tilde{\cal P}^a_{j)}\right]
    \,,
\end{equation}
and
\begin{equation}
    \frac{\partial {\cal T}_{ij}}{\partial {\cal B}_{kl}}
    = \frac{1}{1+\zeta^2} \left[ \delta_{ij} \delta^{kl}
    - \delta_{(i}^k \delta^{l}_{j)}
    + \tensor{\epsilon}{_{(i}^p^{(k}} \tensor{\epsilon}{^{l)}^q_{j)}} v_p v_q \right]
    \,.
\end{equation}
Using the above, we obtain
\begin{widetext}
\begin{eqnarray}
    \!\!&&\!\!
    \int{\rm d}^3 z \left[\frac{\delta {\cal T}_{ij}(x)}{\delta A^r_a(z)} \frac{\delta {\cal T}^{kl}(y)}{\delta {\cal P}^a_r(z)} + \frac{\delta {\cal T}_{ij}(x)}{\delta B^r_a(z)} \frac{\delta {\cal T}^{kl}(y)}{\delta {\cal K}^a_r(z)}\right] \bigg|_{\rm SCS}
    \\
    \!\!&&\!\!\qquad
    = - \frac{1}{1+\zeta^2} \Bigg[ \delta_{r(i} \tilde{\cal K}^a_{j)}
    \left(\delta^{kl} \delta^r_p - \delta_p^{(k} \delta^{l)r} \right) \Gamma_a^p
    + \zeta \delta_{ij} \delta_p^{(k} \delta^{l)r} {\cal K}^a_r K_a^p
    + \left( \delta_{ij} {\cal P}^a_r - \delta_{r(i} \tilde{\cal P}^a_{j)} \right) \delta^{r(k} \delta_q^{l)} K_a^q
    \nonumber\\
    \!\!&&\!\!\qquad\quad
    + \left(v^s \tensor{\epsilon}{_s_{(i}^{(k}}
    - \zeta \delta_{(i}^{(k}\right) \delta_{j)r} \tensor{\epsilon}{^{l)}^p^q} \left({\cal P}^{-1}\right)_c^r {\cal P}^d_p \partial_d {\cal P}_q^c
    \nonumber\\
    \!\!&&\!\!\qquad\quad
    - \frac{\zeta}{1+\zeta^2} \frac{\gamma^2}{2} \left(P_\Psi \left(\zeta - i \gamma^5\right) \left(\gamma^0 - v_u \gamma^u\right) \left[2 \gamma^2 v_{(i} v^{(k}
    + \delta^{(k}_{(i}
    - \gamma^2 \left(\gamma^0 - v_n \gamma^n\right) \gamma_{(i} v^{(k}\right] v_{j)} \gamma^{l)} \Psi
    + c.c.\right) \Bigg] \delta^3(x-y)
    \nonumber\\
    \!\!&&\!\!\qquad\quad
    + \frac{1}{1+\zeta^2} \left[
    v^s \tensor{\epsilon}{_{s}_{(i}^{(k}}
    - \zeta \delta_{(i}^{(k} \right] \tensor{\epsilon}{^{l)}_{j)}^p} {\cal P}^d_p \frac{\partial \delta^3(x-y)}{\partial x^d}
    \,.\nonumber
\end{eqnarray}
\end{widetext}
With this, the bracket (\ref{eq:TT bracket}) follows.

\section{Discrete symmetries}

In this appendix, we gather several useful discrete transformations of fermion bilinears, which can be used to confirm the symmetries discussed in Sec.~(\ref{sec:Discrete symmetries}).

\subsection{Parity}
\begin{eqnarray}
    P^{-1} P_\Psi \Psi P \!\!&=&\!\! P_\Psi \Psi
    \,,\\
    P^{-1} P_\Psi \overline{P_\Psi} P \!\!&=&\!\! P_\Psi \overline{P_\Psi}
    \,,\\
    P^{-1} \overline{\Psi}\Psi P \!\!&=&\!\! \overline{\Psi}\Psi
    \,,
\end{eqnarray}
\begin{eqnarray}
    P^{-1} P_\Psi \gamma^0 {\Psi} P \!\!&=&\!\! P_\Psi \gamma^0 \Psi
    \,,\\
    P^{-1} P_\Psi \gamma^i {\Psi} P \!\!&=&\!\! - P_\Psi \gamma^i \Psi
    \,,
\end{eqnarray}
\begin{eqnarray}
    P^{-1} P_\Psi \gamma^0 \gamma^i {\Psi} P \!\!&=&\!\! - P_\Psi \gamma^0 \gamma^i \Psi
    \,,\\
    P^{-1} P_\Psi \gamma^i \gamma^j {\Psi} P \!\!&=&\!\! P_\Psi \gamma^i \gamma^j \Psi
    \,,
\end{eqnarray}
\begin{eqnarray}
    P^{-1} P_\Psi i S_{0i} {\Psi} P \!\!&=&\!\! - P_\Psi i S_{0i} \Psi
    \,,\\
    P^{-1} P_\Psi i S_i {\Psi} P \!\!&=&\!\! P_\Psi i S_i \Psi
    \,,
\end{eqnarray}
\begin{eqnarray}
    P^{-1} P_\Psi \gamma^0 \gamma^j i S_{0i} {\Psi} P \!\!&=&\!\! P_\Psi \gamma^0 \gamma^j i S_{0i} \Psi
    \,,\\
    P^{-1} P_\Psi \gamma^m \gamma^j i S_{0i} {\Psi} P \!\!&=&\!\! - P_\Psi \gamma^m \gamma^j i S_{0i} \Psi
    \,,
\end{eqnarray}
\begin{eqnarray}
    P^{-1} P_\Psi \gamma^0 \gamma^j i S_i {\Psi} P \!\!&=&\!\! - P_\Psi \gamma^0 \gamma^j i S_i \Psi
    \,,\\
    P^{-1} P_\Psi \gamma^m \gamma^j i S_i {\Psi} P \!\!&=&\!\! P_\Psi \gamma^m \gamma^j i S_i \Psi
    \,,
\end{eqnarray}

\begin{eqnarray}
    P^{-1} P_\Psi i \gamma^5 \Psi P \!\!&=&\!\! - P_\Psi i \gamma^5 \Psi
    \,,\\
    P^{-1} P_\Psi \gamma^i \gamma^0 i \gamma^5 \Psi P \!\!&=&\!\! P_\Psi \gamma^i \gamma^0 i \gamma^5 \Psi
    \,,\\
    P^{-1} P_\Psi i \gamma^5 \gamma^0 \Psi P \!\!&=&\!\! - P_\Psi i \gamma^5 \gamma^0 \Psi
    \,,\\
    P^{-1} P_\Psi i \gamma^5 \gamma^i \Psi P \!\!&=&\!\! P_\Psi i \gamma^5 \gamma^i \Psi
    \,.
\end{eqnarray}

\subsection{Time reversal}
\begin{eqnarray}
    T^{-1} P_\Psi \Psi T \!\!&=&\!\! P_\Psi \Psi
    \,,\\
    T^{-1} P_\Psi\,\overline{P_\Psi} T \!\!&=&\!\! P_\Psi\overline{P_\Psi}
    \,,\\
    T^{-1} \overline{\Psi}\Psi T \!\!&=&\!\! \overline{\Psi}\Psi
    \,,
\end{eqnarray}
\begin{eqnarray}
    T^{-1} P_\Psi \gamma^0 {\Psi} T \!\!&=&\!\! P_\Psi \gamma^0 \Psi
    \,,\\
    T^{-1} P_\Psi \gamma^i {\Psi} T \!\!&=&\!\! - P_\Psi \gamma^i \Psi
    \,,
\end{eqnarray}

\begin{eqnarray}
    T^{-1} P_\Psi \gamma^0 \gamma^i {\Psi} T \!\!&=&\!\! - P_\Psi \gamma^0 \gamma^i \Psi
    \,,\\
    T^{-1} P_\Psi \gamma^i \gamma^j {\Psi} T \!\!&=&\!\! P_\Psi \gamma^i \gamma^j \Psi
    \,,\\
    T^{-1} P_\Psi i S_{0i} {\Psi} T \!\!&=&\!\! - P_\Psi i S_{0i} \Psi
    \,,\\
    T^{-1} P_\Psi i S_i {\Psi} T \!\!&=&\!\! P_\Psi i S_i \Psi
    \,,
\end{eqnarray}
\begin{eqnarray}
    T^{-1} P_\Psi \gamma^0 \gamma^j i S_{0i} {\Psi} T \!\!&=&\!\! P_\Psi \gamma^0 \gamma^j i S_{0i} \Psi
    \,,\\
    T^{-1} P_\Psi \gamma^m \gamma^j i S_{0i} {\Psi} T \!\!&=&\!\! - P_\Psi \gamma^m \gamma^j i S_{0i} \Psi
    \,,
\end{eqnarray}
\begin{eqnarray}
    T^{-1} P_\Psi \gamma^0 \gamma^j i S_i {\Psi} T \!\!&=&\!\! - P_\Psi \gamma^0 \gamma^j i S_i \Psi
    \,,\\
    T^{-1} P_\Psi \gamma^m \gamma^j i S_i {\Psi} T \!\!&=&\!\! P_\Psi \gamma^m \gamma^j i S_i \Psi
    \,,
\end{eqnarray}

\begin{eqnarray}
    T^{-1} P_\Psi i \gamma^5 \Psi T \!\!&=&\!\! - P_\Psi i \gamma^5 \Psi
    \,,\\
    T^{-1} P_\Psi \gamma^i \gamma^0 i \gamma^5 \Psi T \!\!&=&\!\! P_\Psi \gamma^i \gamma^0 i \gamma^5 \Psi
    \,,\\
    T^{-1} P_\Psi i \gamma^5 \gamma^0 \Psi T \!\!&=&\!\! - P_\Psi i \gamma^5 \gamma^0 \Psi
    \,,\\
    T^{-1} P_\Psi i \gamma^5 \gamma^i \Psi T \!\!&=&\!\! P_\Psi i \gamma^5 \gamma^i \Psi
    \,.
\end{eqnarray}

\subsection{Charge conjugation}
\begin{eqnarray}
    C^{-1} P_\Psi \partial_a {\Psi} C \!\!&=&\!\! \partial_a \overline{\Psi}\, \overline{P_\Psi}
    \,,\\
    C^{-1} P_\Psi \overline{P_\Psi} C \!\!&=&\!\! P_\Psi \overline{P_\Psi}
    \,,\\
    C^{-1} \overline{\Psi}\Psi C \!\!&=&\!\! \overline{\Psi}\Psi
    \,,
\end{eqnarray}
\begin{eqnarray}
    C^{-1} P_\Psi \gamma^0 \gamma^i \partial_a {\Psi} C \!\!&=&\!\! \partial_a \overline{\Psi} \gamma^i \gamma^0 \overline{P_\Psi}
    \,,\\
    C^{-1} P_\Psi \gamma^i \gamma^j \partial_a {\Psi} C \!\!&=&\!\! \partial_a \overline{\Psi} \gamma^j \gamma^i \overline{P_\Psi}
    \,,
\end{eqnarray}
\begin{eqnarray}
    C^{-1} P_\Psi i S_{0i} {\Psi} C \!\!&=&\!\! - \overline{\Psi} i S_{0i} \overline{P_\Psi}
    \,,\\
    C^{-1} P_\Psi i S_i {\Psi} C \!\!&=&\!\! - \overline{\Psi} i S_i \overline{P_\Psi}
    \,,
\end{eqnarray}

\begin{eqnarray}
    C^{-1} P_\Psi \gamma^0 \gamma^j i S_{0i} {\Psi} C \!\!&=&\!\! - \overline{\Psi} i S_{0i} \gamma^j \gamma^0 \overline{P_\Psi}
    \,,\\
    C^{-1} P_\Psi \gamma^m \gamma^j i S_{0i} {\Psi} C \!\!&=&\!\! - \overline{\Psi} i S_{0i} \gamma^j \gamma^m \overline{P_\Psi}
    \,,
\end{eqnarray}
\begin{eqnarray}
    C^{-1} P_\Psi \gamma^0 \gamma^j i S_i {\Psi} C \!\!&=&\!\! - \overline{\Psi} i S_i \gamma^j \gamma^0 \overline{P_\Psi}
    \,,\\
    C^{-1} P_\Psi \gamma^m \gamma^j i S_i {\Psi} C \!\!&=&\!\! - \overline{\Psi} i S_i \gamma^j \gamma^m \overline{P_\Psi}
    \,,
\end{eqnarray}

\begin{eqnarray}
    C^{-1} P_\Psi i \gamma^5 \Psi C \!\!&=&\!\! \overline{\Psi} i \gamma^5 \overline{P_\Psi}
    \,,\\
    C^{-1} P_\Psi \gamma^i \gamma^0 i \gamma^5 \Psi C \!\!&=&\!\! \overline{\Psi} i \gamma^5 \gamma^0 \gamma^i \overline{P_\Psi}
    \,,\\
    C^{-1} P_\Psi i \gamma^5 \gamma^I \Psi C \!\!&=&\!\! - \overline{\Psi} \gamma^I i \gamma^5 \overline{P_\Psi}
    \,,\\
    C^{-1} P_\Psi \gamma^I \Psi C \!\!&=&\!\! - \overline{\Psi} \gamma^I \overline{P_\Psi}
    \,.
\end{eqnarray}


\begin{thebibliography}{10}

\bibitem{Ashtekar1}
A. Ashtekar, Phys. Rev. Lett. \textbf{57}, 2244 (1986).

\bibitem{Ashtekar2}
A. Ashtekar, Phys. Rev. D \textbf{36}, 1587 (1987).

\bibitem{Barbero}
J.~F. Barbero, Phys. Rev. D \textbf{51}, 5507 (1995), arXiv:gr-qc/9410014.

\bibitem{Holst}
S. Holst, Phys. Rev. D \textbf{53}, 5966 (1996), arXiv:gr-qc/9511026.

\bibitem{Tetrads}
E.~I. Duque, \textit{Hamiltonian gravity in tetrad-connection variables}, Phys. Rev. D \textbf{(To appear)}, arXiv:2509.06153.

\bibitem{Inconsistency}
J.~C. Baez and K.~V. Krasnov, J. Math. Phys. {\bf 39}, 1251-1271 (1998), arXiv:9703112.

\bibitem{Half-densitized}
T. Thiemann, Class. Quantum Grav. {\bf 15}, 1487 (1998), arXiv:9705021.

\bibitem{Mercuri}
S. Mercuri, Phys. Rev. D \textbf{73}, 084016 (2006).

\bibitem{Perez}
A. Perez and C. Rovelli, Phys. Rev. D \textbf{73}, 044013 (2006).

\bibitem{Freidel}
L. Freidel, D. Minic, and T. Takeuchi, Phys. Rev. D \textbf{72}, 104002 (2005).

\bibitem{Alexandrov}
S. Alexandrov, Class. Quantum Grav. \textbf{25}, 145012 (2008), arXiv:0802.1221.

\bibitem{ADM}
R. Arnowitt, S. Deser, and C.~W. Misner,  in {\em Gravitation: An Introduction to Current Research}, edited by L. Witten (Wiley, New York, 1962), reprinted in \cite{arnowitt2008republication}.

\bibitem{arnowitt2008republication}
R. Arnowitt, S. Deser, and C.~W. Misner, Gen. Rel. Grav. {\bf 40}, 1997 (2008).

\bibitem{pons1997gauge}
J.~M. Pons, D.~C. Salisbury, and L.~C. Shepley, Phys. Rev. D \textbf{658} (1997), arXiv:gr-qc/9612037.

\bibitem{salisbury1983realization}
D.~C. Salisbury, and K. Sundermeyer, Phys. Rev. D {27}, 740 (1983).

\bibitem{EMGCov}
M. Bojowald, and E.~I. Duque, Phys. Rev. D {\bf 108}, 084066 (2023), arXiv:2310.06798.

\bibitem{Parity}
M. Bojowald and R. Das, Phys. Rev. D {\bf 78}, 064009 (2008), arXiv:0710.5722.

\bibitem{Rastgoo}
F. Fragomeno and S. Rastgoo, \textit{Canonical Electrodynamics in Ashtekar-Barbero variables}, arXiv:2507.06276.

\bibitem{hojman1976geometrodynamics}
S.~A. Hojman, K. Kucha\v{r}, and C. Teitelboim, Ann. Phys. (New York) {\bf 96},  88  (1976).

\bibitem{kuchar1974geometrodynamics}
K.~V. Kucha\v{r}, J. Math. Phys. {\bf 15}, 708 (1974).

\bibitem{EMGFT}
E.~I. Duque, \textit{Emergent field theory}, arXiv:2507.16163.

\bibitem{EMGGowdy}
M. Bojowald and E.~I. Duque, Phys. Rev. D, {\bf 110}, 124001, (2024), arXiv:2407.13583.

\bibitem{EMGCosmoK}
M. Bojowald, M. Díaz, E.~I. Duque, \textit{Perturbative emergent modified gravity on cosmological backgrounds: Kinematics}, arXiv:2507.14358.

\bibitem{EMGScalar}
M. Bojowald and E.~I. Duque, Phys. Rev. D 109, 084006, (2024) arxiv:2311.10693.

\bibitem{EMGPF}
E.~I. Duque, Phys. Rev. D {\bf 109}, 044014 (2024), arXiv:2311.08616.



\end{thebibliography}
\end{document}